\documentclass{pasj00}
\color{black}
\begin{document}
\SetRunningHead{M. Komiyama, K. Sato, R. Nagino, T. Ohashi and K. Matsushita}{Suzaku observations of the NGC 5044 group}
\Received{2008/10/31}
\Accepted{}

\title{Suzaku observations of  metallicity distribution in the
intracluster medium of the NGC 5044 group}



%
 \author{%
   Madoka \textsc{Komiyama}\altaffilmark{1},
   Kosuke \textsc{Sato}\altaffilmark{1,2},
   Ryo \textsc{Nagino}\altaffilmark{1},
   Takaya \textsc{Ohashi}\altaffilmark{3}
  \\ and
   Kyoko \textsc{Matsushita}\altaffilmark{1}}
 \altaffiltext{1}{Department of Physics, Tokyo University of Science,
                  1-3 Kagurazaka, Shinjyuku-ku, Tokyo 162-8601}
 \email{j1207629@ed.kagu.tus.ac.jp}
 \altaffiltext{2}{Graduate School of Natural Science and Technology,
                  Kanazawa University,
		  Kakuma, Kanazawa, Ishikawa 920-1192}
 \altaffiltext{3}{Department of Physics, Tokyo Metropolitan University,
                  1-1 Minami-Osawa, Hachioji, Tokyo 192-0397}

\KeyWords{galaxies:abundances --- clusters of galaxies:intracluster medium ---
 groups:individual (NGC 5044 group)} 

\maketitle
\begin{abstract}
The
metallicity distribution in the intracluster medium of the NGC 5044
group was studied up to $0.3 r_{180}$ using the XIS instrument on board
the Suzaku satellite.
Abundances of O, Mg, Si, S, and Fe were measured with high accuracy.
The region within a radius of $0.05 r_{180}$ from the center shows
approximately solar abundances of Mg, Si, S, and Fe, while the O/Fe ratio
is about 0.5--0.6 in solar units.  In the outer region, the Fe abundance
gradually drops to 0.3 solar.  Radial abundance profiles of Mg, Si
and S are similar to that of Fe, while that of O seems to be flatter.
At $r>0.05 r_{180}$, the mass density profile of O differs from that
of Fe, showing a shoulder-like structure that traces the luminosity
density profile of galaxies.  The mass-to-light ratios for O and Fe in
NGC 5044 are one of the largest among groups of galaxies, but they are
still smaller than those in rich clusters. These abundance features
probably reflect the metal enrichment history of this relaxed group
hosting a giant elliptical galaxy in the center.
\end{abstract}

\section{Introduction}
\label{sec:intro}
Groups and clusters of galaxies include 
most  galaxies in the nearby universe.
\citep{Mulchaey2000}. 
These systems also represent building blocks of
rich clusters and are  the best laboratories for the
study of thermal and chemical history governed by baryons.  An
important clue in studying the evolution of galaxies is the
abundances of different elements in the hot X-ray emitting gas in
groups or clusters of galaxies: namely the intracluster medium
(ICM)\@.  The metals in the ICM have been synthesized by supernova (SN) in
galaxies.  As a result, ratios of ICM metal mass to the total light
from galaxies in clusters or groups, i.e.,\ metal-mass-to-light ratios,
are key parameters in investigating the chemical evolution of the
ICM\@.

Since Fe lines are prominent in X-ray spectra, the content and
distribution of Fe in the ICM have been studied in detail.  The ASCA
satellite first enabled us to measure the distribution of Fe in the
ICM\@ (e.g., Fukazawa et al.\ 2000; Finoguenov et al.\ 1999).  Poor
clusters and groups of galaxies exhibit different properties from
richer systems in that the derived iron-mass-to light ratios (IMLR)
are systematically smaller \citep{Makishima2001}.

\citet{Rasmussen2007} reported the temperature and abundance
 profiles of 15 nearby  groups of galaxies observed by Chandra.
They showed that Fe abundances of the groups declined
 from about 1 solar at the center
 to about 0.1 solar at $r_{500}$, which is about $0.54~r_{180}$.
Si shows less systematic radial variation, indicating 
SN II contribution increases with radius.
\citet{Finoguenov2007} also reported the metal distributions of groups
of galaxies observed by XMM-Newton.
They also found that metallicity of some groups drop to 0.1 solar
at $\sim r_{500}$.

O and Mg are synthesized predominantly  in SN II, while Fe and Si are
synthesized in both SN Ia and SN II\@.  Abundance measurements
spanning the range of species from O to Fe are therefore required for
unambiguous determination of the formation history of massive stars.
XMM-Newton provided the means to constrain O and Mg abundances in some
systems (e.g.,\ \cite{Matsushita2003},\cite{Tamura2003},
\cite{Matsushita2007b}), however, reliable results have been obtained
only for the central regions of very bright clusters or groups of
galaxies dominated by cD galaxies.

Suzaku (\cite{Mitsuda2007}) is the fifth Japanese X-ray astronomy
satellite.  The X-ray Imaging Spectrometer(XIS) instrument (\cite{Koyama2007}) offers an improved
line spread function due to a very small low-pulse-height tail in the
energy range below 1 keV coupled with a very low background.
Therefore, especially for the regions of low surface brightness or
equivalent width, the XIS provides better sensitivity for O lines.
The instrumental Al line of the MOS detectors
 on XMM-Newton causes problems in
 measuring the Mg abundance in somewhat fainter systems.

With the Suzaku satellite, the oxygen-mass-to-light ratio (OMLR) as
well as the IMLR of several clusters of galaxies and several groups of
galaxies were measured up to 0.2--0.3$r_{180}$
(\cite{Matsushita2007a}, \cite{Tokoi2008}, \cite{kSato2007},
\cite{kSato2008a}, \cite{kSato2008b}, \cite{kSato2008c}) .  The OMLR
and IMLR increase with the radius, and become relatively flat beyond
0.1 $r_{180}$.
The Fornax cluster, which is the nearest poor cluster with asymmetric
X-ray morphology, shows the smallest IMLR and OMLR up to 0.13
$r_{180}$ \citep{Matsushita2007a}.

The NGC~5044 group is a nearby group of galaxies whose redshift is
$z=0.009020$ from the NASA/IPAC Extragalactic Database (NED).  The group
has a giant elliptical galaxy NGC~5044 at the center.  The X-ray
emission of the group shows nearly symmetric spatial distribution
\citep{David1994}.
The ICM properties in this group have been studied well using
XMM-Newton and Chandra by \citet{Buote2003a} and \citet{Buote2003b},
and the metal distribution in the outer regions was studied
using XMM-Newton \citep{Buote2004}.
They revealed that the Fe abundance of the ICM drops
 from about $1$ solar within $r\approx 50~{\rm kpc}$
 to $\sim 0.4$ solar near $r\approx 100~{\rm kpc}$
 using the solar abundances by \citet{grsa1998}
\citep{Buote2003b}.
In the outer region, at $\sim$ 176-352~${\rm kpc}$, 
 the iron abundance is determined to be about $0.15$ solar
 using the same solar abundances \citep{Buote2004}.
\citet{Tamura2003} showed O abundance to be $0.25$ solar within the
central $10-20~{\rm kpc}$ using the Reflection Grating Spectrometers (RGS)
on  XMM-Newton.  The mass distribution has been studied using
Chandra and ROSAT data (\cite{David1994}, \cite{Betoya-Nonesa2006}).

We use the Hubble constant $H_{0} = 70~{\rm km/s/Mpc}$.  The distance to
the NGC~5044 group is $D_{\rm L} = 38.9~{\rm Mpc}$, and $1'$
corresponds to $11.1~{\rm kpc}$.  The virial radius,
$r_{180}=1.95~(H_0/100)^{-1} \sqrt{k \langle T \rangle /10~{\rm
keV}}~{\rm Mpc}$ (\cite{Markevitch1998}, \cite{Evrard1996}), is about
$880~{\rm kpc}$ for the average temperature $k \langle T \rangle =
1.0~{\rm keV}$.  We use the new abundance table from \citet{lodd2003}
in this paper.  Abundances of O and Fe are about $1.7$ times
and $1.6$ times higher than those of  \citet{angr1989}, respectively.
Unless otherwise specified, errors are quoted at 90\% confidence for the
single parameter of interest.

\section{Observations}
\label{sec:obs}
\begin{figure}
  \begin{center}
    \FigureFile(80mm,80mm){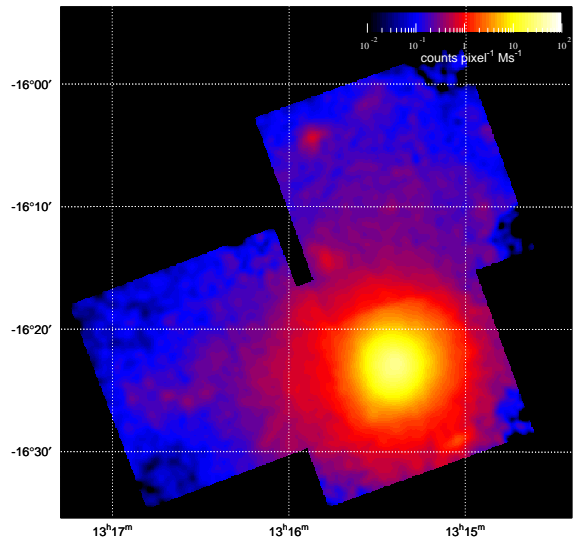}
  \end{center}
 \caption{ XIS image (0.5--4.0 keV) of the NGC~5044 Group. 
 The NXB (non X-ray background) and
 CXB (cosmic X-ray background) were subtracted,
 and the difference in  exposure times was corrected.
The image was smoothed with a Gaussian of $\sigma=16~{\rm pixels}$.
}
  \label{fig:n5044_from_ksato}
\end{figure}

Suzaku performed three pointing observations of the NGC~5044 group
in July 2006.  The observational log is shown in Table
\ref{tab:obsinfo}.  The first observation (hereafter, the central
field) had a pointing direction toward the cD galaxy, NGC~5044.  The
second and  third observations were centered $15'$ north and $15'$
east of NGC~5044, respectively (hereafter, the north and east fields,
respectively).  Figure \ref{fig:n5044_from_ksato} shows a
0.5-$4.0~{\rm keV}$ image for the three fields.  The observed region
covers a distance of about $24~{\rm arcmin}$, or $\sim 267~{\rm kpc}$,
from NGC~5044.

We also analyzed the MOS and PN data from the XMM-Newton observation of
NGC~5044.  Details of the data reduction are given in
\citet{Nagino2008}.

\begin{longtable}{lllll}
\caption{Suzaku observations of the NGC 5044 group}
\label{tab:obsinfo}
  \hline              
Fields & Seq. No. & (RA, Dec) in J2000& Date of obs. & Exp. time \\
\endfirsthead
  \hline
\endlastfoot
  \hline
Center & 801046010
       & (\timeform{13h15m24s.0},~\timeform{-16D23'07''.8})
       & 2006/07/02 & $19.7$ ks \\
North  & 801047010
       & (\timeform{13h15m24s.0},~\timeform{-16D08'07''.8})
       & 2006/07/03 & $54.6$ ks \\
East   & 801048010
       & (\timeform{13h16m26s.5},~\timeform{-16D23'07''.8})
       & 2006/07/04 & $62.4$ ks \\
\end{longtable}

\section{Spectral Analysis}
\label{sec:ana}
The  XIS was operated
in the nominal mode.  We included the data formats of both $5\times5$ and
$3\times3$ editing modes in our analysis using {\tt
XSELECT(Ver.~2.3)}.  The analysis was performed using {\tt
HEAsoft(Ver.~6.0.6)} and {\tt XSPEC(Ver.~11.3.2o)}.  After applying
the standard data selection criteria, the exposure times of center,
north, and east fields were $19.7,~54.6,$ and $62.4~{\rm ks}$,
respectively.

The response of the XRT and XIS was calculated using the {\tt xissimarfgen}
ancillary response file (ARF) generator \citep{Ishisaki2007} and {\tt
xisrmfgen} response matrix file (RMF) generator,  version
{\tt 2006-10-26}.  Slight degradation of the energy resolution was
considered in the RMF, and decrease in the low-energy transmission of
the XIS optical blocking filter (OBF) was included in the ARF\@.  The
ARF response was calculated assuming a surface brightness profile,
$S(r)$, based on the analysis of XMM-Newton data by \citet{Nagino2008}.

For the central field, we analyzed the spectra extracted from four
annular regions of 0-2',~2-4',~4-6', and 6-9' centered on
NGC~5044.  For the north and east fields, we used the spectra
extracted from two annular regions of $r<15'$ and $r>15'$ centered on
the galaxy.
Each spectrum was binned
 to  observe details in metal lines, especially O lines.
Each spectral  bin contained 50 or more counts.

The non X-ray background (NXB) was subtracted from the spectra using
the database of  night Earth observations with Suzaku for the same
detector area and with the same distribution of {\it COR} (Tawa et
al. 2007).  We produced the spectra for an energy range 0.4--7.0$~{\rm
keV}$, since above $7.0~{\rm keV}$, subtraction of  background
lines is difficult.  The spectra of XIS detectors (XIS0, XIS1, XIS2,
XIS3) were fitted simultaneously with a model consisting of the ICM
emission with additional power-law and the Galactic components.

\subsection{The ICM component}
\label{sec:ana_fit}
First, we assumed the ICM consisting of a single temperature {\it vapec}
\citep{Smith2001} model (hereafter, the 1T model).  The metal abundances
of He, C, N, and Al were fixed to the solar values.  We divided
the other metals into seven groups, O; Ne; Mg; Si; S, Ar and Ca; Fe; and
Ni, and allowed them to vary.  The spectra were subject to a common
interstellar absorption, $N_{\rm{H}}$, fixed at the Galactic values,
which are $5.03 \times 10^{20},~ 4.88 \times 10^{20},~ 5.21 \times
10^{20} ~{\rm (atoms/cm^2)}$ for the center, north, and east fields,
respectively.

We also applied a two-temperature (hereafter, 2T) {\it vapec}
model for the ICM, where the metal abundances of the two components
were assumed to have the same value.

\subsection{CXB and discrete sources}
\label{sec:ana_CXBNXB}
The cosmic X-ray background (CXB) was modeled by a power-law spectrum
with a photon index $\Gamma = 1.4$.  The normalization was allowed to
vary, which can also account for contributions from discrete sources in
galaxies and a possible hard emission.

\subsection{The Galactic component}
\label{sec:ana_Galactic}
The Galactic emission mainly arises from the local hot bubble (LHB)
and the Milky-Way halo (MWH). We used two-temperature {\it apec}
thermal spectra for this emission, and by combining it with the 1T model
for the ICM we fitted the observed spectra for the outermost
($r>15'$) regions in the north and  east fields.
\color{black} The temperature and normalization of the two components
in the Galactic emission were left free with metal abundance fixed to
the solar level.
The results of the spectral fits are shown in Table \ref{tab:Galkt}.
The temperature values of the two apec models are consistent with the
typical values for the LHB and MWH derived by XMM \citep{Lumb2002}.

Since the errors for these temperatures are small, we fixed the
temperatures of the LHB and MWH to $0.12~{\rm keV}$ and $0.30~{\rm
keV}$, respectively.  Assuming that the Galactic component has the
same surface brightness within the entire observed region, we
calculated the normalization of the Galactic component in each region
based on the surface brightness measured in the outermost region in
the north field.
 The spectra of the other  regions were 
consistently reproduced  with the  
sum of the ICM model and  this Galactic component.
This feature supports the view that these  {\it apec} components
 in the outermost region are dominated by the emission in our Galaxy.

\begin{table}
\caption{Temperature of the Galactic components with the 1T model for
 the ICM.}
\label{tab:Galkt}
\begin{center}
\begin{tabular}{lcccc}
\hline
Region & $kT_{\rm LHB}$\footnotemark[$\ast$]
       & $kT_{\rm MWH}$\footnotemark[$\ast$]
       & Ratio of
       & $\chi^2$/d.o.f. \\
(arcmin/$r_{180}$)
 & (keV) & (keV)
 & $Norm$\footnotemark[$\dagger$]
 & - \\ \hline
$> 15 / 0.19$ (N)
           & $0.12^{+0.02}_{-0.01}$
           & $0.30^{+0.03}_{-0.03}$
           & $0.48^{+0.21}_{-0.21}$
           & $1113/895$\\
$> 15 / 0.19$ (E)
           & $0.13^{+0.02}_{-0.01}$
           & $0.29^{+0.07}_{-0.02}$
           & $0.47^{+0.19}_{-0.23}$
           & $996/895$\\
\hline
\multicolumn{5}{{@{}l@{}}}{\hbox to 0pt{\parbox{85mm}{\footnotesize
\par\noindent
\footnotemark[$\dagger$]
 Ratio of normalization $Norm_{\rm MWH}/Norm_{\rm LHB}$
\par\noindent
\footnotemark[$\ast$]
 LHB = Local Hot Bubble,~MWH = Milky Way Halo
}\hss}}
\end{tabular}
\end{center}
\end{table}

\section{Results}
\label{sec:res}
\subsection{Fitting results}
\label{sec:res_fit}
Table \ref{tab:ktno} summarizes the best-fit parameters of
temperature, $\chi^2$ values, degrees of freedom, and abundances.  For
the 2T fitting, the ratios of the {\it vapec} normalizations and flux
between cool and hot ICM components are also summarized in Table
\ref{tab:ktno}.

Within 
$r=0.050~r_{180}$, or $4'$,
 the 2T model significantly improved the fit.  For
$0.050~r_{180} < r < 0.19~r_{180}$ ($4'<r<15'$),
 the 2T model provided a slightly better fit than the 1T model,
while at 
$r> 0.19~r_{180}~(15')$
 the $\chi^2$ values derived for the 1T and 2T models
were nearly equal.

\begin{figure*}
  \begin{center}
    \FigureFile(80mm,60mm){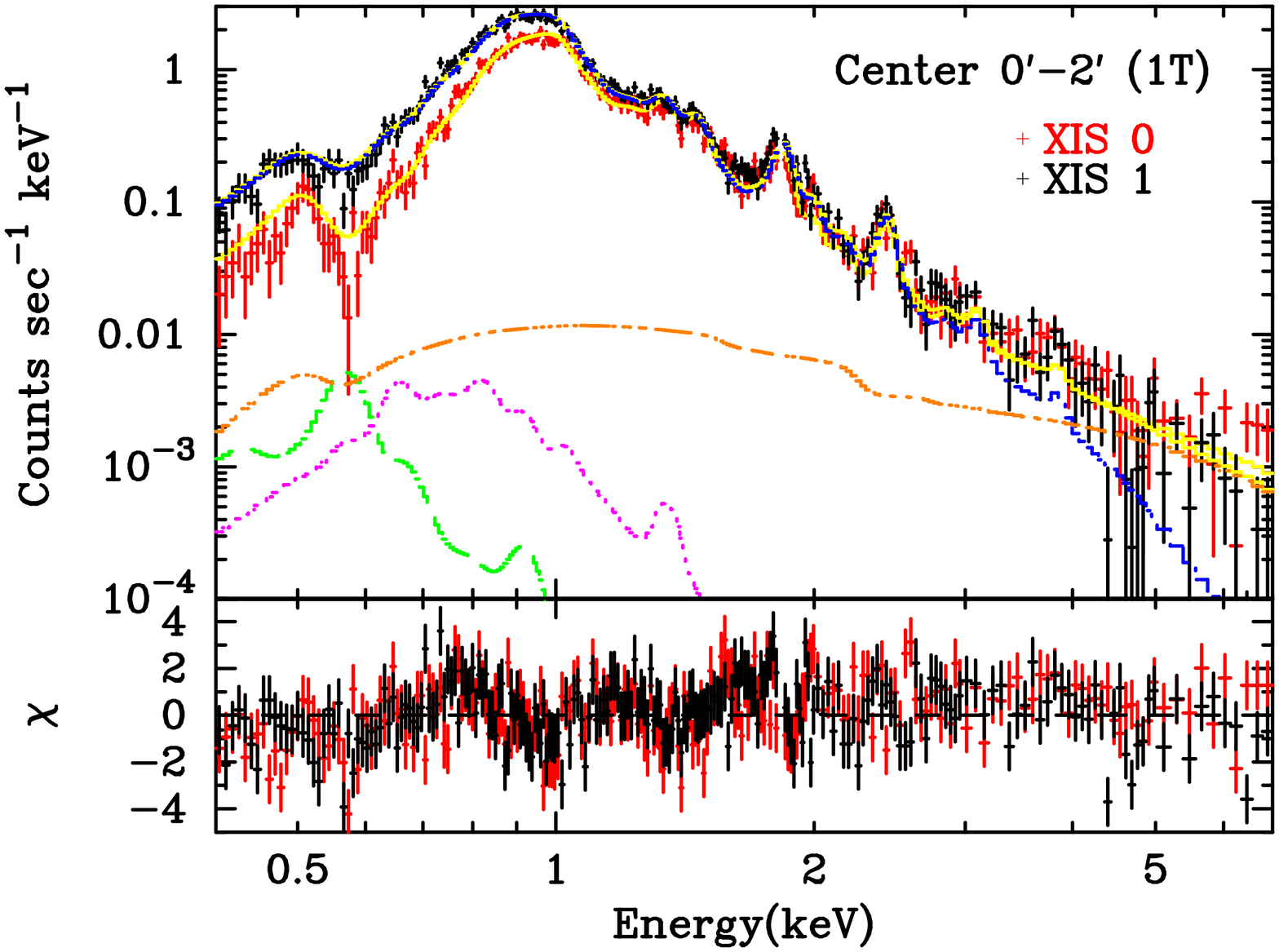}
    \FigureFile(80mm,60mm){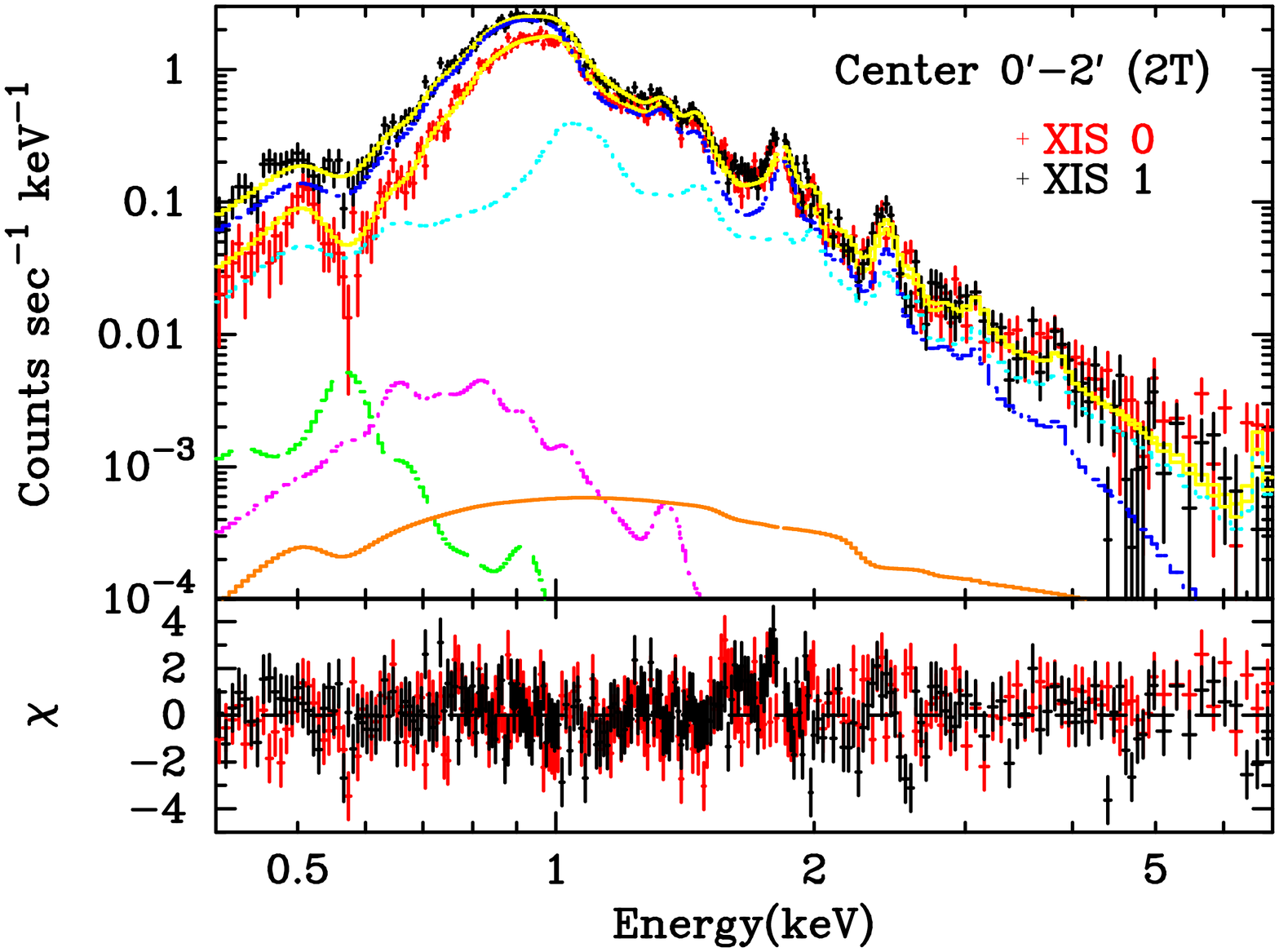}
    \FigureFile(80mm,60mm){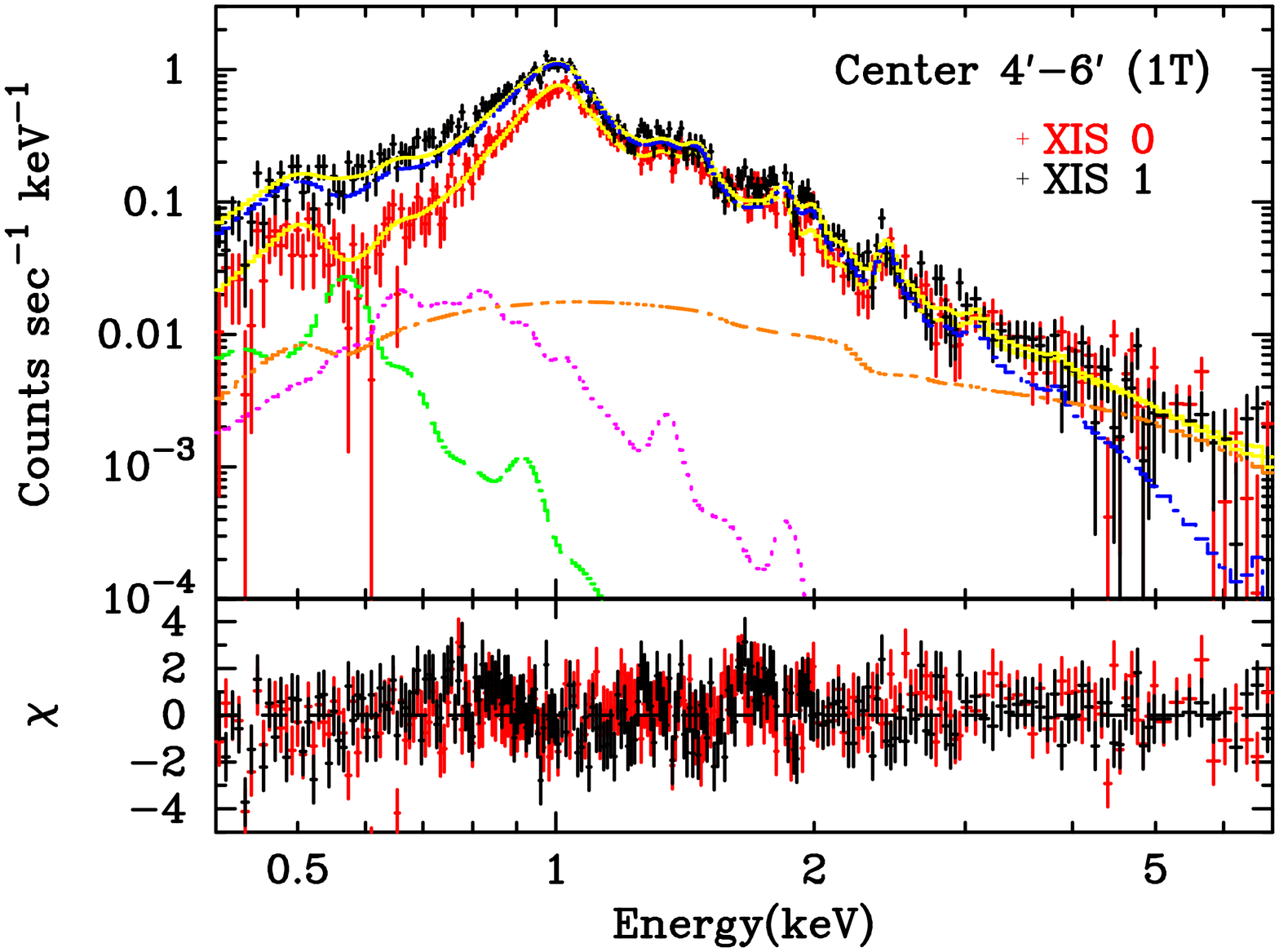}
    \FigureFile(80mm,60mm){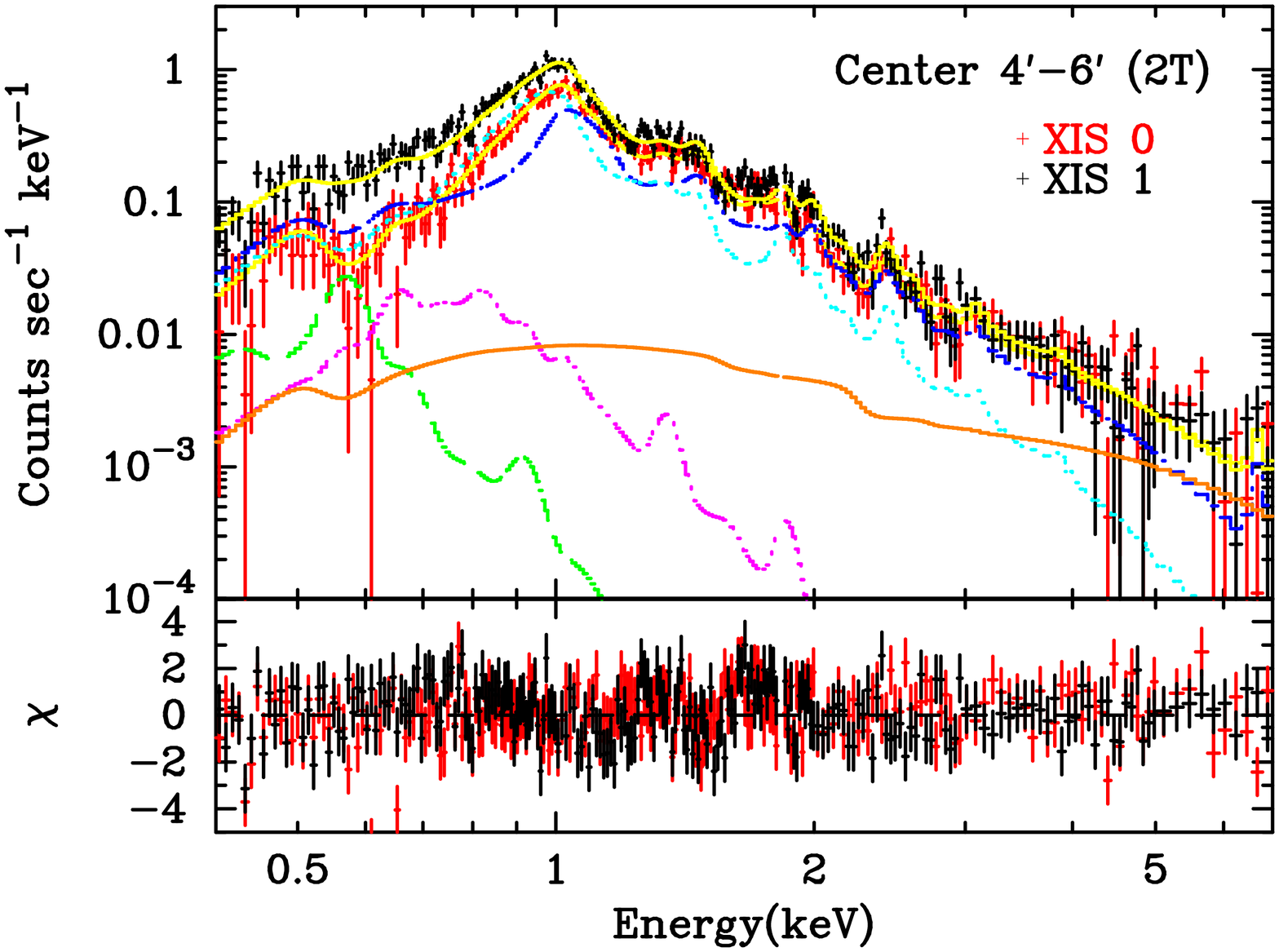}
    \FigureFile(80mm,60mm){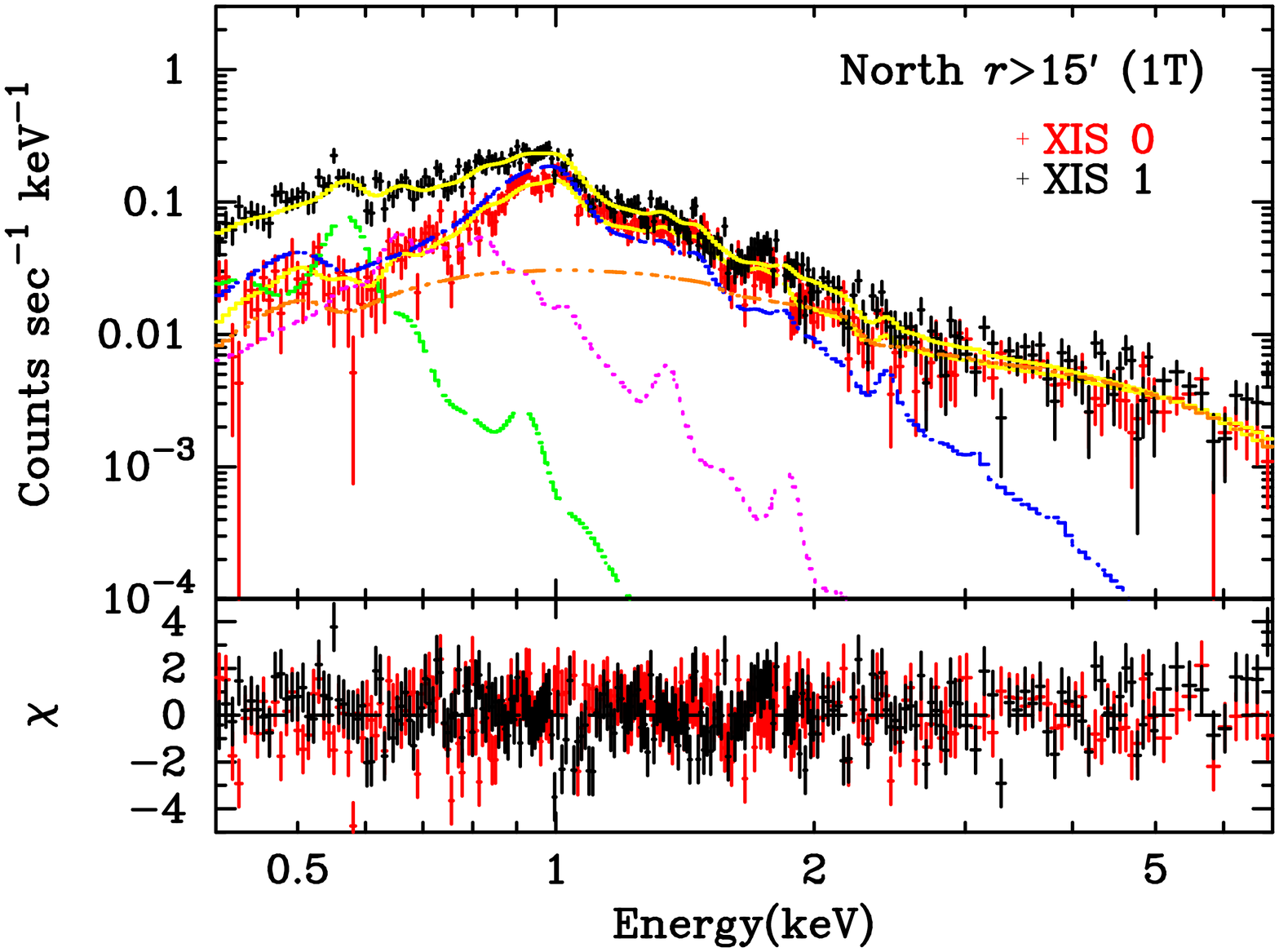}
    \FigureFile(80mm,60mm){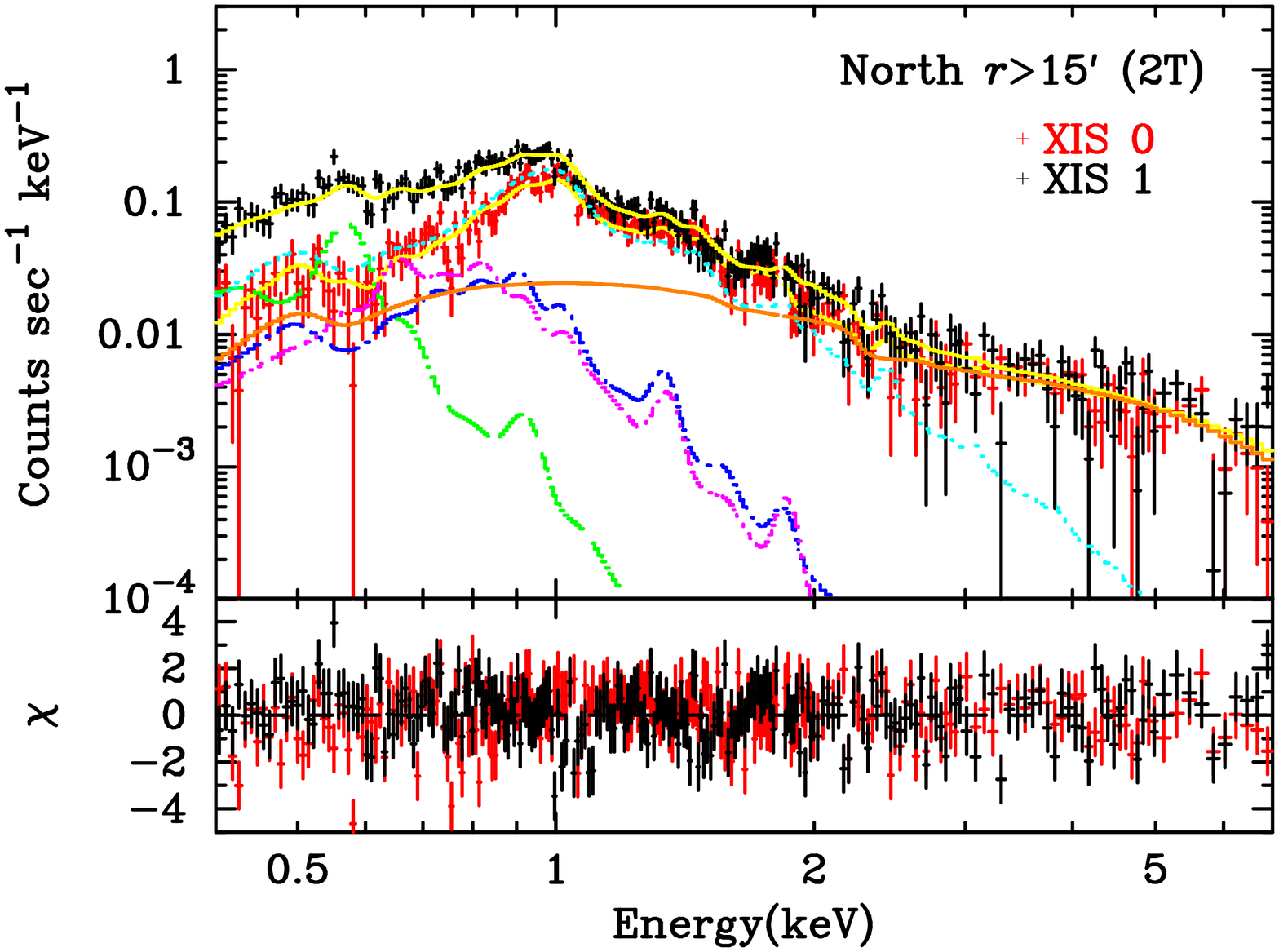}
  \end{center}
  \caption{Representative spectra from the XIS-0 (red) and the XIS-1
 (black) instruments fitted with the 1T (blue) or 2T (blue and light
 blue) models for the ICM including the power-law (orange) and the
 Galactic components (magenta and green). The lower panels show the
 fit residuals in units of $\sigma$.  }
\label{fig:1T2Tspec}
\end{figure*}

In Figure \ref{fig:1T2Tspec}, representative spectra obtained using the
XIS-0 and XIS-1 instruments are shown.  The spectra are well fitted
with the 2T model, and the 1T model provides reasonable fits in the outer
region.

\begin{longtable}{lccccccc}
\caption{
The ICM temperature, $\chi^2$, the ratio of the normalization
and abundances of elements
derived from the 
spectral fits  for the NGC 5044 group. Errors are in the 90\% confidence
range of statistical errors, and do not include systematic errors.}
\label{tab:ktno}
  \hline
\endfirsthead
\multicolumn{8}{{@{}l@{}}}{\hbox to 0pt{\parbox{180mm}{\footnotesize
\par\noindent
\footnotemark[$\ast$]
 Ratio of flux in the best fit model
 ${\rm Flux}_{\rm cool}/{\rm Flux}_{\rm hot}$.
\par\noindent
\footnotemark[$\dagger$]
 Ratio of normalization $Norm_{\rm cool}/Norm_{\rm hot}$.
}\hss}}
\endlastfoot
Fitting & \multicolumn{2}{l}{1T model for the ICM}
 & \multicolumn{5}{l}{2T model for the ICM} \\
Region & $kT$ & $\chi^2$/d.o.f.
 & $kT_{\rm cool}$ & $kT_{\rm hot}$ & Ratio of
 & Ratio of & $\chi^2$/d.o.f. \\
(arcmin/$r_{180}$)
 & (keV) & -
 & (keV) & (keV) & $Norm$\footnotemark[$\dagger$]
 & Flux\footnotemark[$\ast$] & - \\ \hline
$0 - 2 / 0 - 0.025$
 & $0.795^{+0.006}_{-0.006}$ & $1574/899$
 & $0.779^{+0.007}_{-0.007}$ & $1.57^{+0.10}_{-0.06}$
 & $2.59^{+0.34}_{-0.33}$ & $3.88$ & $1276/897$ \\
$2 - 4 / 0.025 - 0.050$
 & $0.96^{+0.01}_{-0.01}$ & $1441/899$
 & $0.827^{+0.007}_{-0.008}$ & $1.41^{+0.06}_{-0.08}$	 
 & $0.88^{+0.11}_{-0.09}$ & $1.25$ & $1206/897$ \\
$4 - 6 / 0.050 - 0.076$
 & $1.05^{+0.01}_{-0.01}$ & $1194/899$
 & $0.96^{+0.05}_{-0.12}$ & $1.48^{+0.18}_{-0.18}$
 & $0.72^{+0.30}_{-0.26}$ & $0.90$ & $1123/897$ \\
$6 - 9 / 0.076 - 0.11$
 & $1.06^{+0.02}_{-0.02}$ & $1027/899$
 & $1.01^{+0.03}_{-0.08}$ & $1.74^{+0.28}_{-0.21}$
 & $0.77^{+0.26}_{-0.27}$ & $0.94$ & $967/897$ \\
 \hline
$< 15 / 0.19$ (N)
 & $1.05^{+0.02}_{-0.02}$ & $1047/899$
 & $1.01^{+0.03}_{-0.04}$ & $2.14^{+0.64}_{-0.41}$
 & $1.30^{+0.43}_{-0.40}$ & $1.32$ & $999.5/897$ \\
$< 15 / 0.19$ (E)
 & $1.06^{+0.02}_{-0.02}$ & $963/899$
 & $1.02^{+0.02}_{-0.05}$ & $1.72^{+0.39}_{-0.24}$
 & $1.44^{+0.60}_{-0.51}$ & $1.61$ & $927/897$ \\
$> 15 / 0.19$ (N)
 & $0.90^{+0.04}_{-0.05}$ & $1091/899$
 & $0.24^{+0.02}_{-0.02}$ & $0.95^{+0.04}_{-0.04}$
 & $0.89^{+0.22}_{-0.22}$ & $0.27$ & $1112/897$ \\
$> 15 / 0.19$ (E)
 & $0.97^{+0.05}_{-0.07}$ & $970/899$
 & $0.91^{+0.09}_{-0.07}$ & $1.77^{+78.13}_{-1.57}$
 & $3.05^{+3.40}_{(-3.05)}$ & $2.95$ & $968/897$ \\
\hline\hline
\multicolumn{8}{c}{1T model for the ICM} \\ \hline
Region & O & Ne & Mg & Si & S & Fe & Ni \\
(arcmin/$r_{180}$)
 & (solar) & (solar) & (solar) & (solar)
 & (solar) & (solar) & (solar) \\ \hline
$0 - 2 / 0 - 0.025$
         & $0.29^{+0.10}_{-0.09}$	 & $0.86^{+0.22}_{-0.21}$
	 & $0.94^{+0.09}_{-0.08}$	 & $0.70^{+0.07}_{-0.06}$
	 & $1.22^{+0.15}_{-0.14}$	 & $0.65^{+0.04}_{-0.04}$
	 & $3.10^{+0.34}_{-0.31}$ \\
$2 - 4 / 0.025 - 0.050$
         & $0.49^{+0.13}_{-0.12}$	 & $0.63^{+0.29}_{-0.27}$
	 & $0.86^{+0.09}_{-0.08}$	 & $0.59^{+0.06}_{-0.06}$
	 & $0.95^{+0.12}_{-0.11}$	 & $0.62^{+0.05}_{-0.05}$
	 & $2.35^{+0.32}_{-0.30}$ \\
$4 - 6 / 0.050 - 0.076$
         & $0.15^{+0.16}_{-0.15}$	 & $0.76^{+0.44}_{-0.43}$
	 & $0.69^{+0.12}_{-0.11}$	 & $0.47^{+0.07}_{-0.07}$
	 & $0.71^{+0.13}_{-0.12}$	 & $0.48^{+0.05}_{-0.05}$
	 & $1.64^{+0.36}_{-0.32}$ \\
$6 - 9 / 0.076 - 0.11$
         & $0.30^{+0.18}_{-0.17}$	 & $0.89^{+0.44}_{-0.44}$
	 & $0.69^{+0.13}_{-0.12}$	 & $0.37^{+0.07}_{-0.07}$
	 & $0.52^{+0.14}_{-0.13}$	 & $0.35^{+0.04}_{-0.04}$
	 & $1.53^{+0.37}_{-0.32}$ \\ \hline
$< 15 / 0.19$ (N)
         & $0.32^{+0.17}_{-0.15}$	 & $0.65^{+0.33}_{-0.34}$
	 & $0.47^{+0.10}_{-0.09}$	 & $0.24^{+0.06}_{-0.05}$
	 & $0.43^{+0.11}_{-0.11}$	 & $0.27^{+0.04}_{-0.04}$
	 & $0.98^{+0.28}_{-0.25}$ \\
$< 15 / 0.19$ (E)
         & $0.13^{+0.16}_{-0.13}$	 & $0.91^{+0.38}_{-0.38}$
	 & $0.60^{+0.11}_{-0.11}$	 & $0.39^{+0.07}_{-0.06}$
	 & $0.56^{+0.13}_{-0.12}$	 & $0.34^{+0.04}_{-0.04}$
	 & $1.35^{+0.33}_{-0.29}$ \\
$> 15 / 0.19 $ (N)
         & $< 0.12$	                 & $0.59^{+0.29}_{-0.30}$
	 & $0.38^{+0.11}_{-0.10}$	 & $0.15^{+0.07}_{-0.07}$
	 & $0.35^{+0.20}_{-0.19}$	 & $0.22^{+0.06}_{-0.05}$
	 & $0.99^{+0.36}_{-0.35}$ \\
$> 15 / 0.19$ (E)
         & $< 0.32$                      & $0.45^{+0.48}_{-0.45}$
	 & $0.37^{+0.19}_{-0.16}$	 & $0.12^{+0.11}_{-0.10}$
	 & $0.43^{+0.30}_{-0.27}$	 & $0.28^{+0.09}_{-0.08}$
	 & $0.83^{+0.65}_{-0.55}$ \\
\hline\hline
\multicolumn{8}{c}{2T model for the ICM} \\ \hline
Region & O & Ne & Mg & Si & S & Fe & Ni \\
(arcmin/$r_{180}$)
 & (solar) & (solar) & (solar) & (solar)
 & (solar) & (solar) & (solar) \\ \hline
$0 - 2 / 0 - 0.025$
         & $0.58^{+0.17}_{-0.15}$	 & $1.77^{+0.43}_{-0.35}$
	 & $1.51^{+0.18}_{-0.16}$	 & $1.00^{+0.12}_{-0.11}$
	 & $1.31^{+0.19}_{-0.17}$	 & $1.25^{+0.13}_{-0.12}$
	 & $2.89^{+0.85}_{-0.68}$ \\
$2 - 4 / 0.025 - 0.050$
         & $0.76^{+0.21}_{-0.19}$	 & $2.24^{+0.56}_{-0.55}$
	 & $1.51^{+0.20}_{-0.18}$	 & $0.92^{+0.11}_{-0.11}$
	 & $1.10^{+0.17}_{-0.14}$	 & $1.24^{+0.13}_{-0.12}$
	 & $1.68^{+0.59}_{-0.51}$ \\
$4 - 6 / 0.050 - 0.076$
         & $0.28^{+0.23}_{-0.20}$	 & $1.62^{+0.36}_{-0.33}$
	 & $0.98^{+0.20}_{-0.18}$	 & $0.60^{+0.11}_{-0.10}$
	 & $0.75^{+0.15}_{-0.14}$	 & $0.87^{+0.11}_{-0.10}$
	 & $1.00^{+0.54}_{-0.46}$ \\
$6 - 9 / 0.076 - 0.11$
         & $0.53^{+0.29}_{-0.25}$	 & $1.80^{+0.80}_{-0.72}$
	 & $0.98^{+0.23}_{-0.20}$	 & $0.47^{+0.12}_{-0.10}$
	 & $0.48^{+0.16}_{-0.15}$	 & $0.78^{+0.12}_{-0.12}$
	 & $0.82^{+0.63}_{-0.53}$ \\ \hline
$< 15 / 0.19$ (N)
         & $0.61^{+0.27}_{-0.24}$	 & $1.31^{+0.58}_{-0.52}$
	 & $0.67^{+0.18}_{-0.15}$	 & $0.29^{+0.09}_{-0.08}$
	 & $0.44^{+0.15}_{-0.14}$	 & $0.49^{+0.11}_{-0.11}$
	 & $1.26^{+0.68}_{-0.60}$ \\
$< 15 / 0.19$ (E)
         & $0.27^{+0.23}_{-0.20}$	 & $1.47^{+0.59}_{-0.51}$
	 & $0.80^{+0.19}_{-0.15}$	 & $0.50^{+0.10}_{-0.09}$
	 & $0.58^{+0.15}_{-0.14}$	 & $0.57^{+0.11}_{-0.09}$
	 & $1.30^{+0.61}_{-0.47}$ \\
$> 15 / 0.19$ (N)
         & $0.18^{+0.05}_{-0.04}$	 & $0.36^{+0.22}_{-0.25}$
	 & $0.44^{+0.18}_{-0.14}$	 & $0.17^{+0.10}_{-0.08}$
	 & $0.34^{+0.22}_{-0.20}$	 & $0.29^{+0.07}_{-0.07}$
	 & $0.88^{+0.38}_{-0.44}$ \\
$> 15 / 0.19$ (E)
         & $0.04^{+0.47}_{-0.04}$	 & $0.82^{+0.83}_{-0.81}$
	 & $0.48^{+0.33}_{-0.22}$	 & $0.14^{+0.16}_{-0.13}$
	 & $0.49^{+0.40}_{-0.32}$	 & $0.34^{+0.19}_{-0.14}$
	 & $1.20^{+1.54}_{-0.96}$ \\
\hline
\end{longtable}

\subsection{Temperature profile of the ICM}
\label{sec:res_kt}
\begin{figure*}
  \begin{center}
    \FigureFile(80mm,60mm){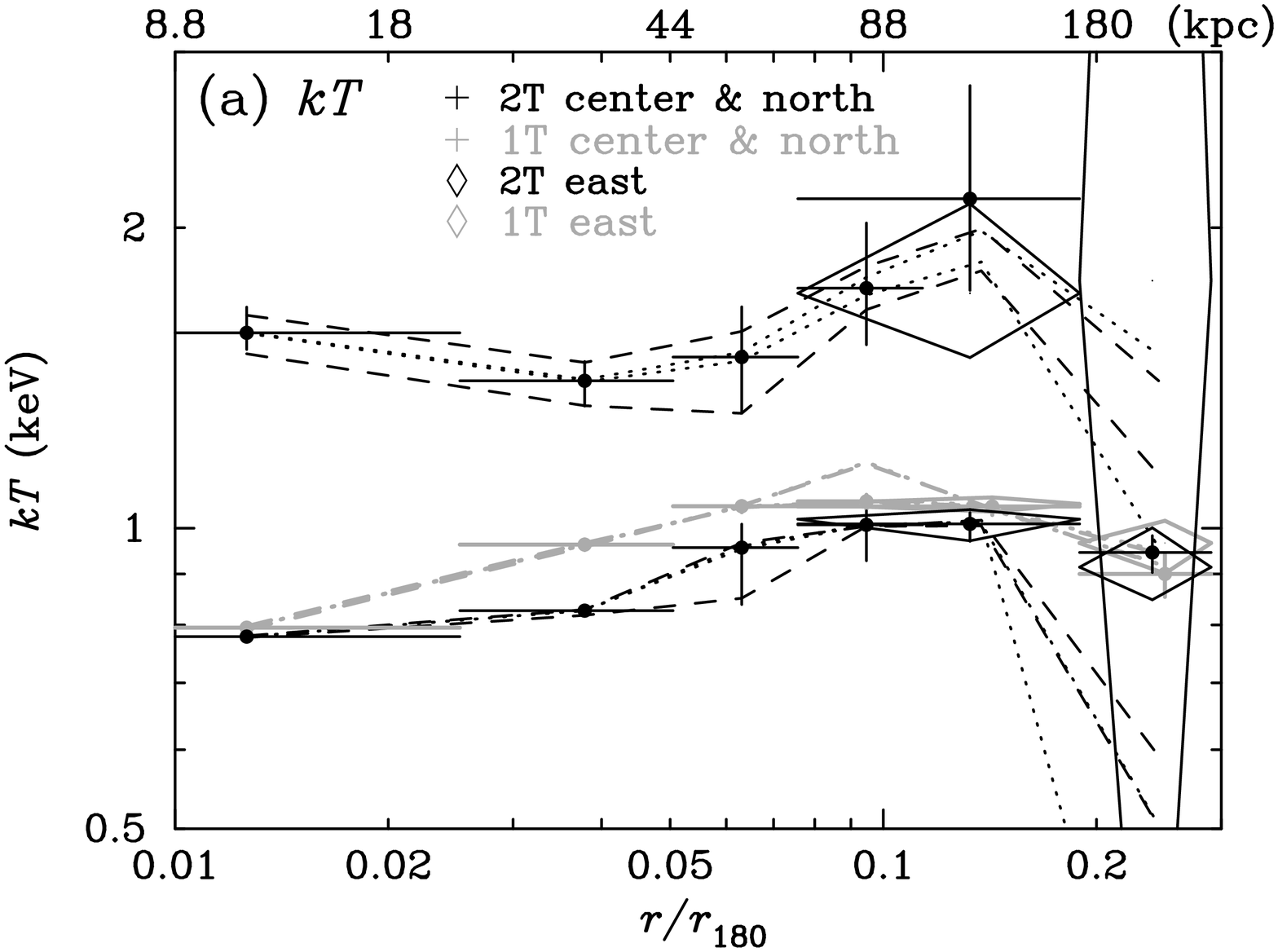}
    \FigureFile(80mm,60mm){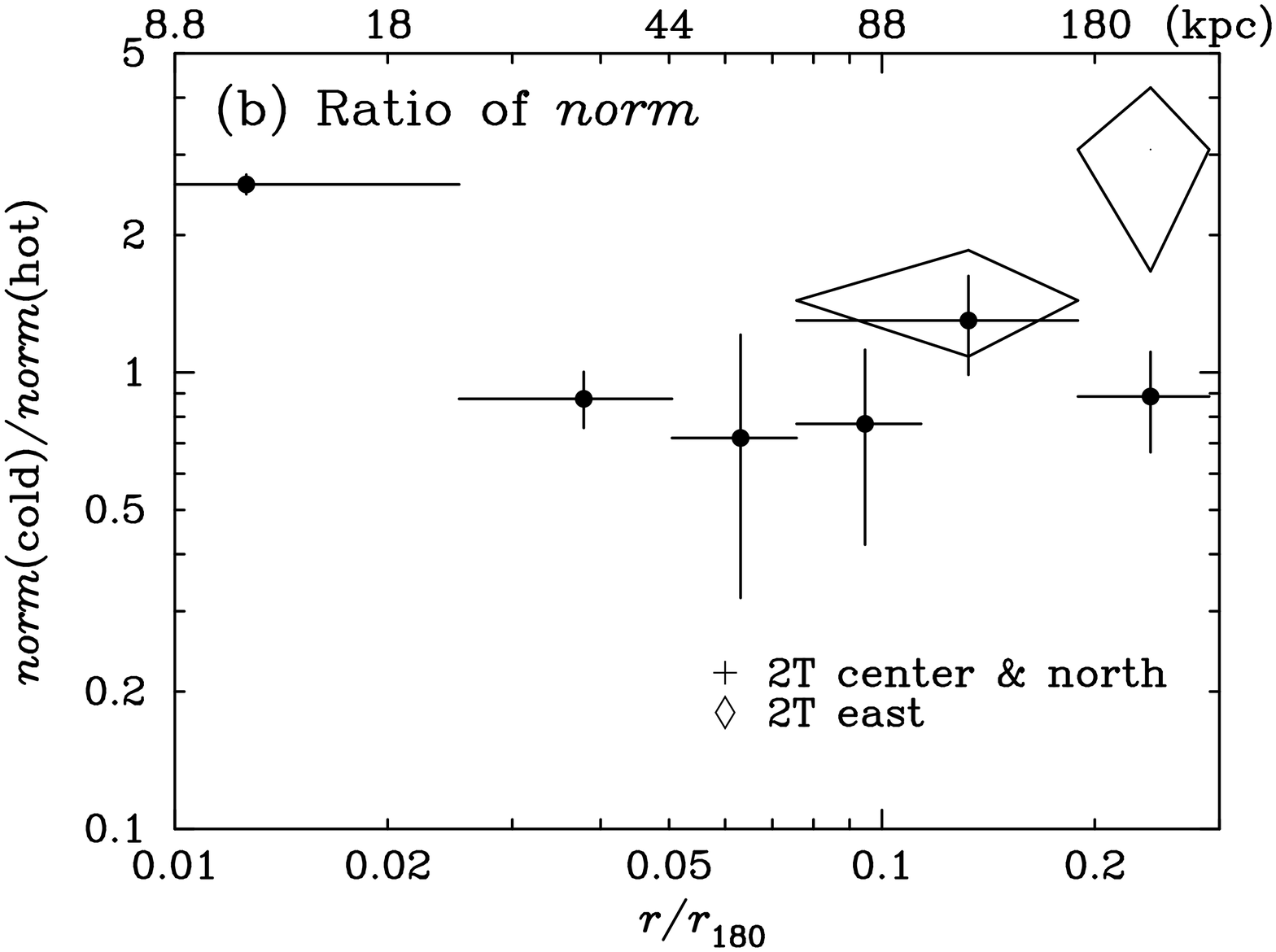}
  \end{center}
  \caption{
 {\bf (a)}
 Temperature profile of the ICM derived with
the 1T (gray) and the 2T (black) models for 
the center and the north fields (crosses) and for the east field
(diamonds), respectively.
 Dotted lines indicate shifts of the best-fit values
 when the NXB level is changed by $\pm 10\%$.
The dashed lines denote shifts
 when the thickness of the OBF contaminant is changed by $\pm 10\%$.
 {\bf (b)} The normalization ratio of the cool and hot
components derived from the 2T model fit.}
\label{fig:kt_sys}
\end{figure*}

We derived the radial temperature profile based on the spectral fit
for the annular regions from the group center to $r \sim 0.3~r_{180}$
(Figure \ref{fig:kt_sys}a).
The fit with the 1T model, within $r<0.1~r_{180}$, showed a
temperature increase from 0.8 keV to 1.1 keV\@.  At $r>0.2~r_{180}$,
it  dropped slightly to 1 keV\@.

For the 2T model, the temperature of one component showed  similar
features to that for the 1T fit.  Within $r<0.1\sim0.2~r_{180}$, we
required another component with $kT\sim 1.5$ keV\@.
The normalization ratio for the cool and  hot {\it vapec}
components is shown in Figure \ref{fig:kt_sys} (b).  The cool
component is the strongest in the innermost region.
In the outer region, the two components show similar normalizations.

\begin{figure*}
  \begin{center}
    \FigureFile(50mm,40mm){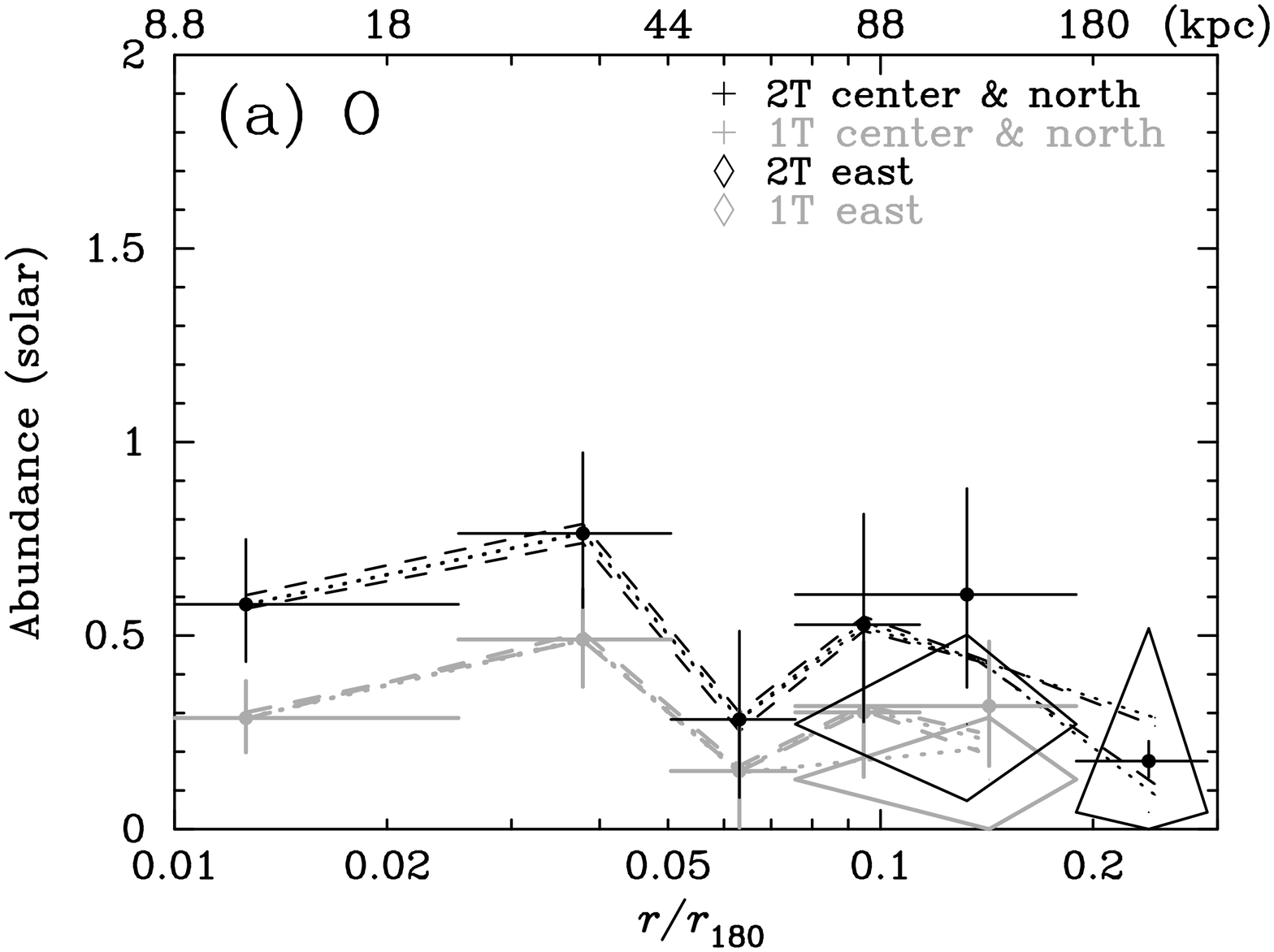}
    \FigureFile(50mm,40mm){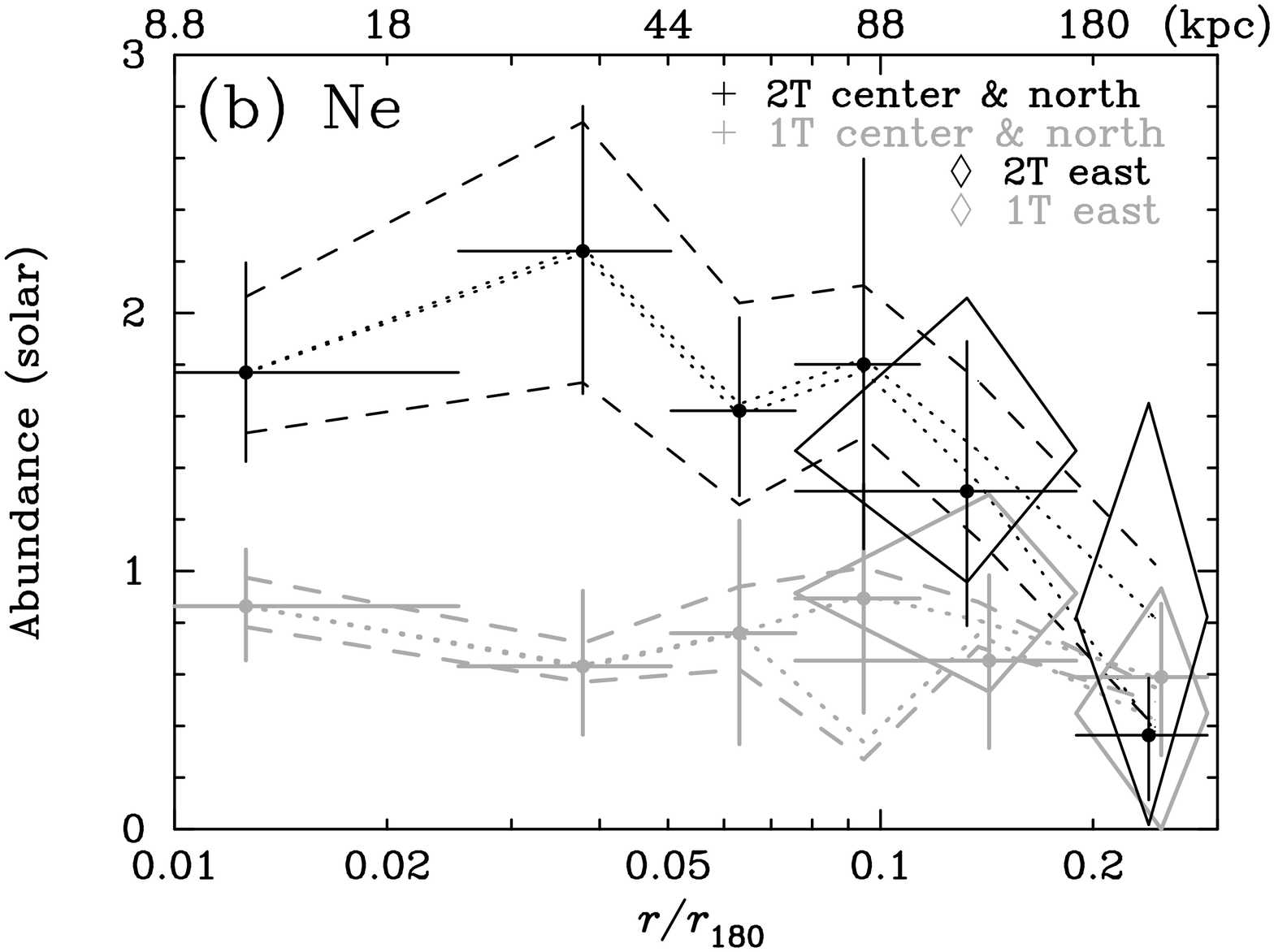}
    \FigureFile(50mm,40mm){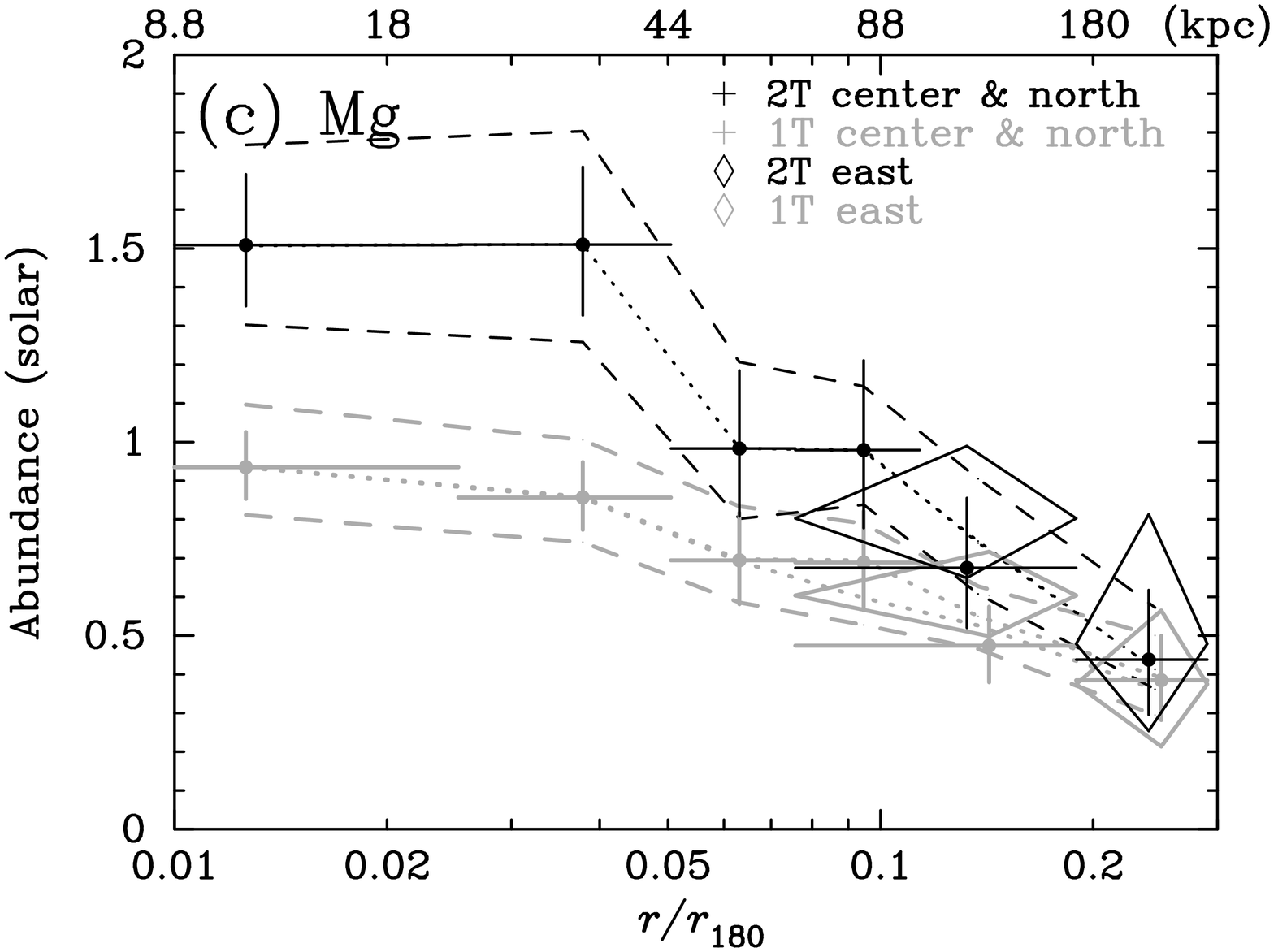}
    \FigureFile(50mm,40mm){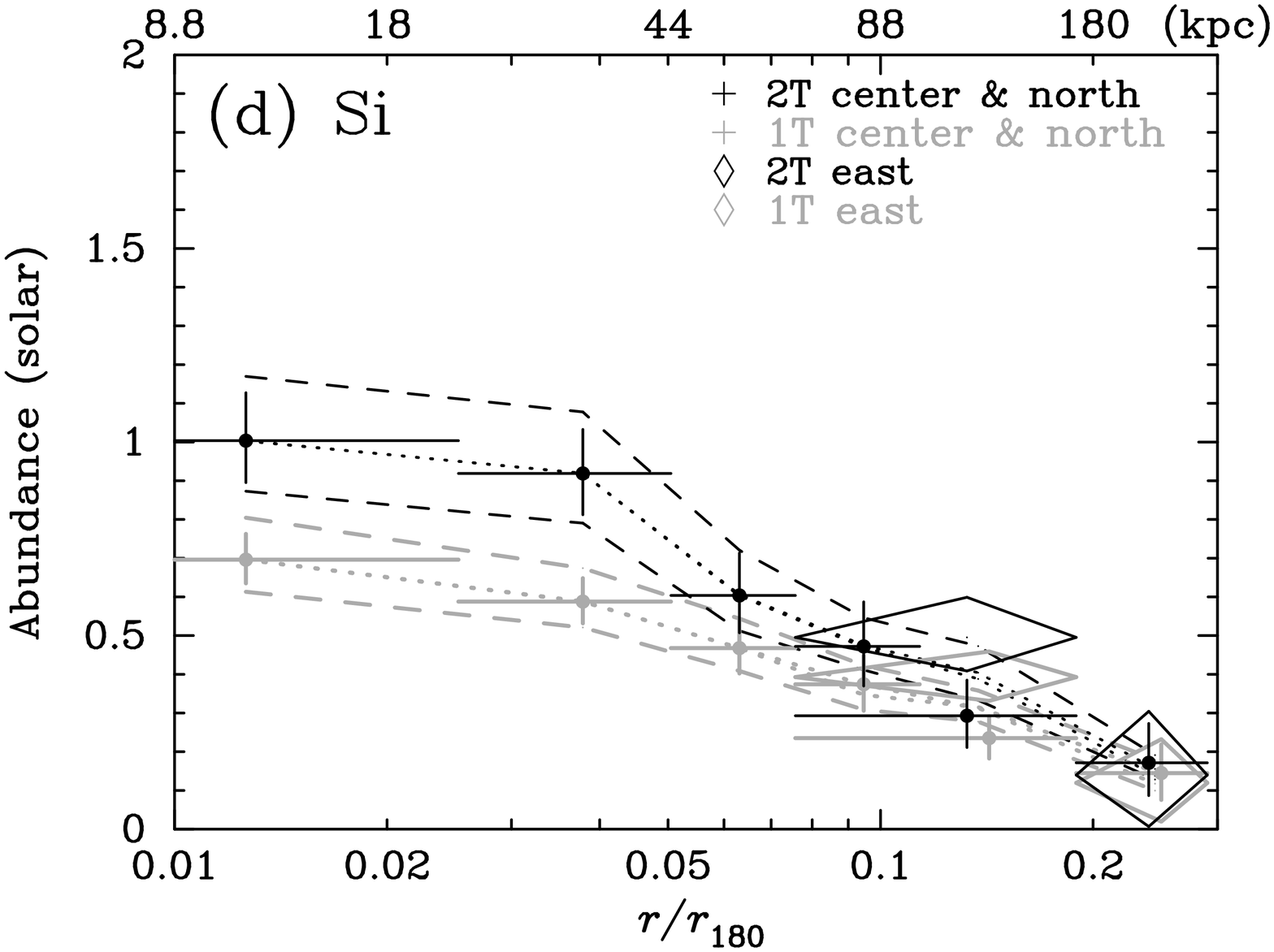}
    \FigureFile(50mm,40mm){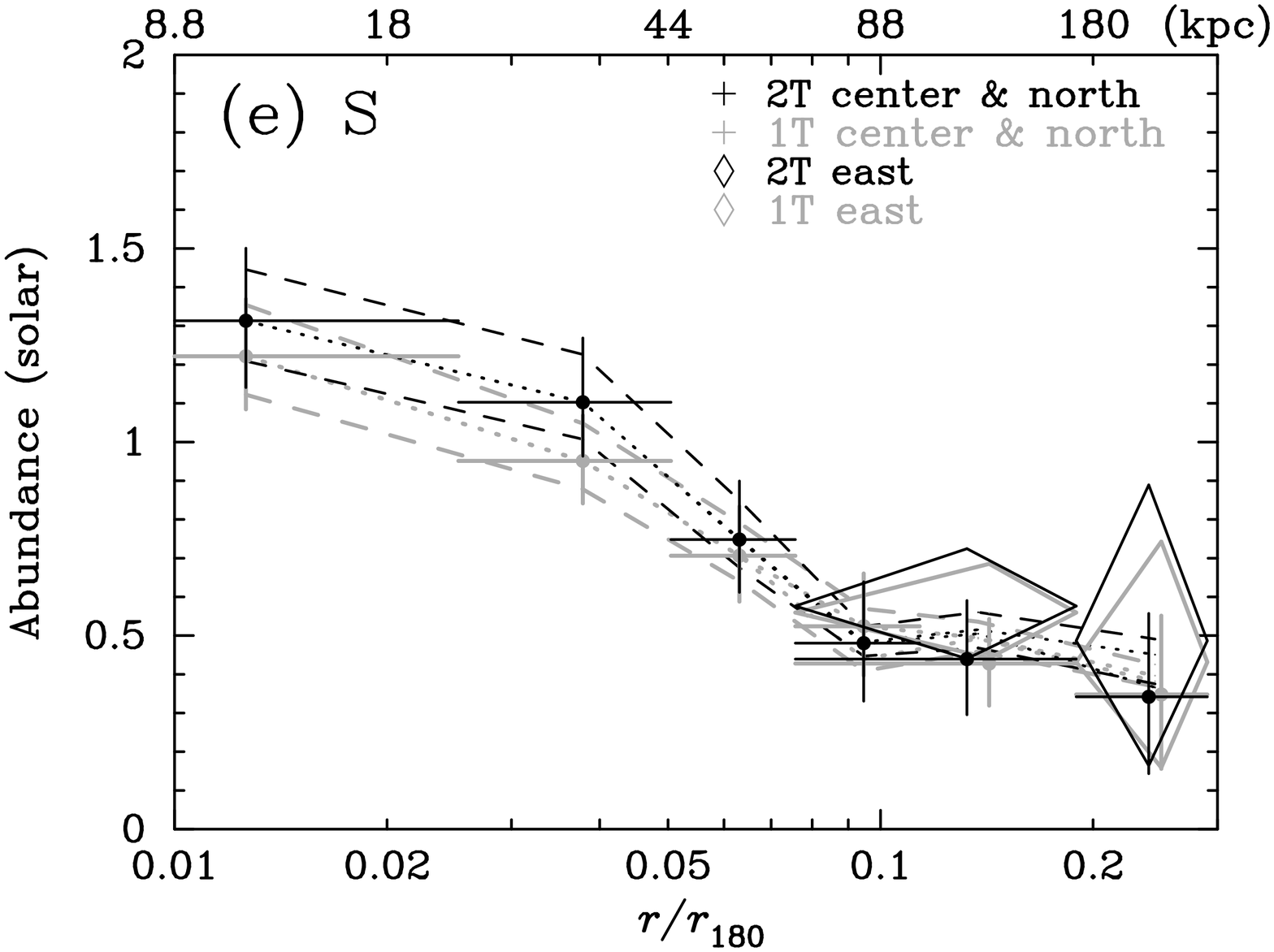}
    \FigureFile(50mm,40mm){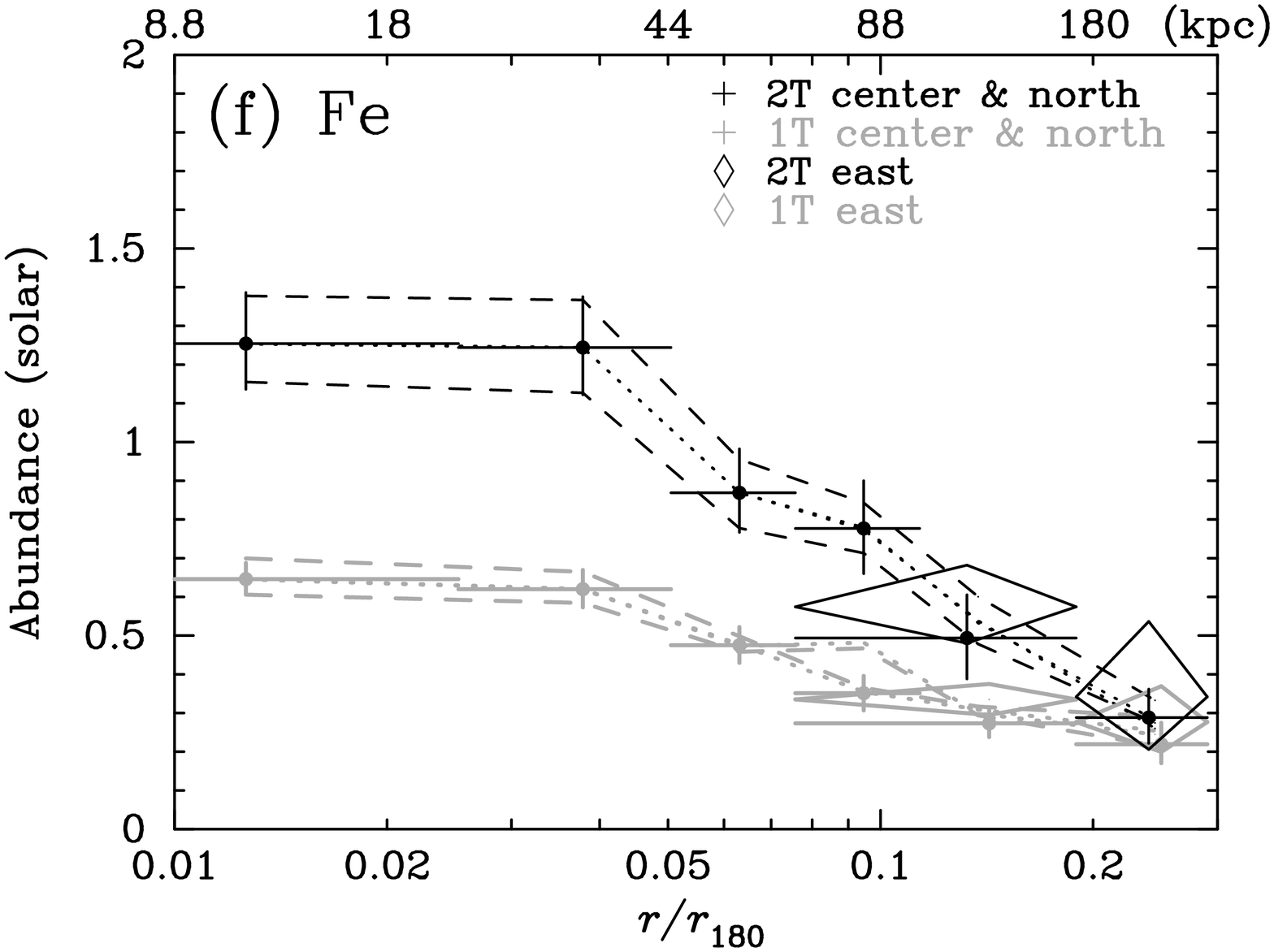}
  \end{center}
  \caption{ Abundance profiles of (a) O, (b) Ne, (c) Mg, (d) Si, (e) S
and (f) Fe of the ICM derived from the 1T  (gray) and  2T
(black) models for the center and  north fields (crosses) and for 
east field (diamonds), respectively.
 The dotted lines indicate shifts of the best-fit
values when the NXB level is changed by $\pm 10\%$.  The dashed lines
denote shifts when the thickness of the OBF contaminant is changed by
$\pm 10\%$.}
\label{fig:abund_sys}
\end{figure*}

\subsection{Abundance profiles of the ICM}
\label{sec:res_abund}
Radial profiles up to  $r < 0.3~r_{180}$, or $260~{\rm kpc}$,
 of  abundances 
of O, Ne, Mg, Si, S, Ar and Ca and Fe are summarized
 in Table \ref{tab:ktno} and Figure \ref{fig:abund_sys}.

On the whole, abundances are high at the center,
 and decrease toward the outer region.
At the center, the 1T and  2T models give an Fe abundance to be
 $\sim 0.7~{\rm solar}$ and $\sim 1.3~{\rm solar}$, respectively.
These abundances decrease to about $0.3~{\rm solar}$ at $r \sim
 0.3~r_{180}$.  In the outer region, the north and  east fields
 give similar Fe abundances suggesting that the metals are mixed over
 a scale of 100 kpc
, or $\sim 0.1~r_{180}$.

The abundance profiles of the other elements are shown
in Figure \ref{fig:abund_sys} (b)-(f).
The O profile looks almost flat, although the statistical errors are
fairly large.  The 1T model fit gives the O abundance to be $\sim$ 0.3
solar, and the 2T model fit gives a two times that value.

The abundance profiles of Mg, Si and S also show negative gradients
and resembles the Fe profile. All these abundances  drop to about
1/2--1/3 of their central values.  The 2T model gives higher Mg and
Si abundances in the central regions than the 1T model, while the S
abundance shows little dependence on the temperature model.  Again, Mg,
Si, and S abundances in the north and east fields are consistent
with each other.

The Ne and Ni abundances are significantly higher than those of other
elements.  Since the K-shell lines of Ne and L-shell lines of Ni are
completely mixed with the Fe-L lines, the derived abundances of Ne and
Ni might include fairly large systematic uncertainties.

\citet{Buote2004} reported that the Fe abundance is about $0.15$ solar at
0.2-0.4$~r_{\rm vir}$ using the solar abundances by \citet{grsa1998},
 where the solar Fe abundance relative to H is  3.16$\times 10^{-5}$
by number.
In contrast, we have derived about $0.3$ solar at $\sim 0.2-0.3~r_{180}$
based on the solar abundances of \citet{lodd2003},
 where the solar Fe abundance relative to H is 2.95$\times 10^{-5}$ by number.
 There is some
difference in modeling the background (e.g., the Galactic emission),
and the best-fit elemental abundance ratios in our analysis differs from
that in \citet{Buote2004}. This may cause the discrepancy in the Fe
abundances. 
When  the
spectra at $r>15'$ were fitted with the 1T  ICM model without 
 the Galactic emission, we obtained the Fe abundance of 0.1--0.2 solar.
However, this model were not able to reproduce the observed spectra
at lower energy band, and  $\chi^2$ increased significantly.
Another possibility is that at the outer region of
0.3--0.4$r_{180}$ the Fe abundance might be very much lower than 0.3
solar.  \color{black}

\subsection{Profiles of abundance ratios of the ICM}
\begin{figure*}
  \begin{center}
    \FigureFile(160mm,160mm){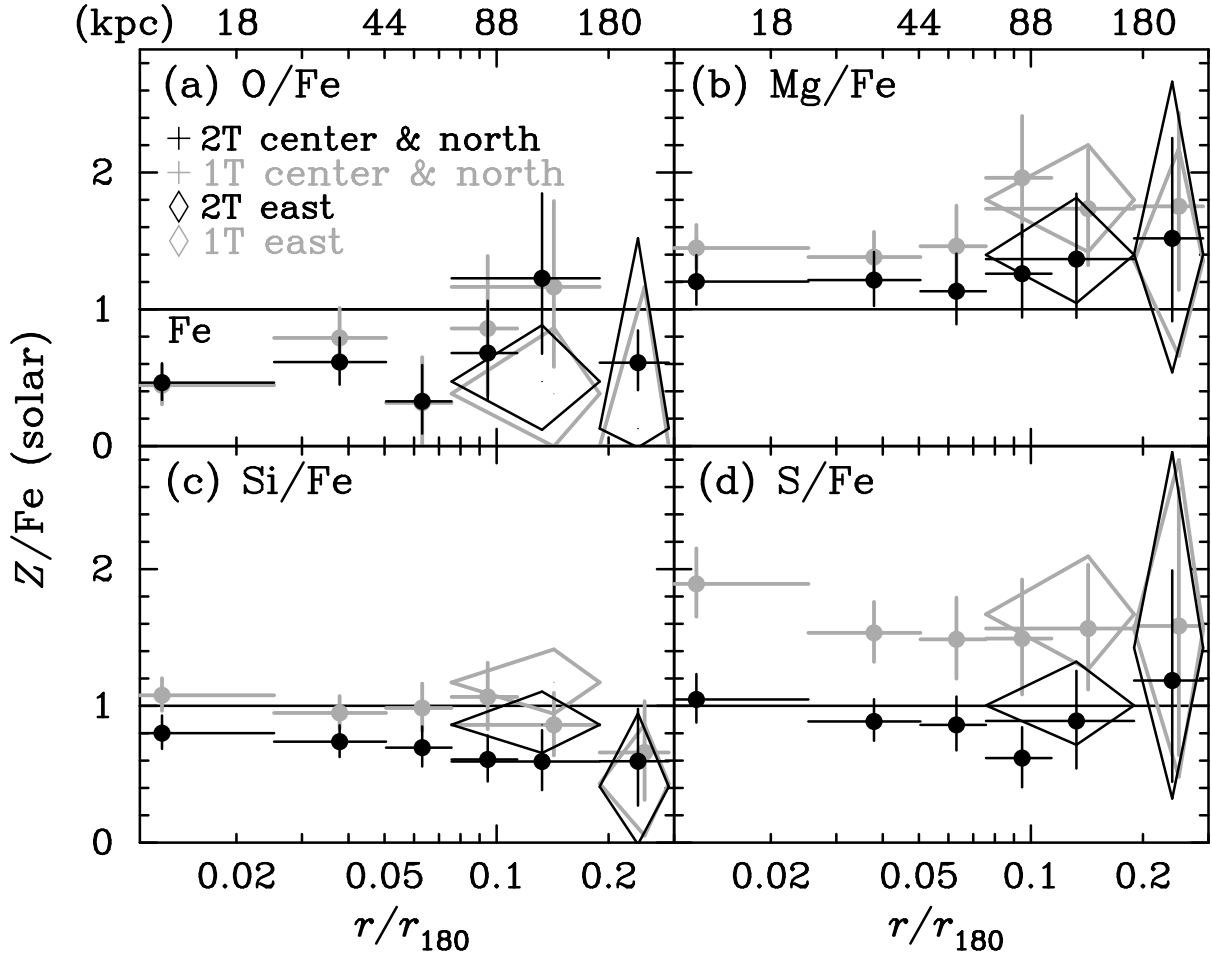}
  \end{center}
  \caption{Radial profiles of the abundance ratios of O, Mg, Si, and S
 divided by the Fe abundance (in units of solar ratio). The meanings
of colors and symbols are the same as in Figure \ref{fig:abund_sys}.}
  \label{fig:abundfe}
\end{figure*}

Figure \ref{fig:abundfe} summarizes the abundance ratios of O, Mg, Si,
and S divided by the Fe value, all taken in solar units.
The abundance ratios, Mg/Fe, and  S/Fe, are consistent with the
absence of
radial gradients.
The O/Fe ratio is also consistent
with having no radial gradient within rather large error bars.
The discrepancies between the 1T and 2T models become much smaller,
except for the S/Fe ratio.

The weighted average of the abundance ratios is calculated for the
central and offset regions and
summarized in Table \ref{tab:aveabund}.
On average,
 the abundance ratios are mostly consistent between the
center and  offset regions.
The 2T model indicates O/Fe, Mg/Fe, Si/Fe, and S/Fe abundance ratios
to be 0.6, 1.3, 0.7 and 1, respectively.  The Ne/Fe ratio might have
additional systematic uncertainty due to the overlap with  Fe-L
lines.

\begin{longtable}{cccccc}
\caption{Weighted averages of the abundance ratios in  solar units.
}
\label{tab:aveabund}
  \hline              
\endfirsthead
  \hline
\endlastfoot
\multicolumn{6}{c}{Single temperature fitting} \\ \hline
field & O/Fe & Ne/Fe & Mg/Fe & Si/Fe & S/Fe \\ \hline
center   & $0.60^{+0.17}_{-0.16}$	 & $1.62^{+0.43}_{-0.42}$
	 & $1.56^{+0.15}_{-0.14}$	 & $1.02^{+0.09}_{-0.08}$
	 & $1.60^{+0.16}_{-0.15}$ \\
offset   & $0.77^{+0.36}_{-0.35}$	 & $2.36^{+0.73}_{-0.71}$
	 & $1.66^{+0.31}_{-0.27}$	 & $0.78^{+0.17}_{-0.15}$
	 & $1.59^{+0.43}_{-0.39}$ \\
all      & $0.69^{+0.22}_{-0.19}$	 & $1.99^{+0.42}_{-0.41}$
	 & $1.61^{+0.17}_{-0.15}$	 & $0.90^{+0.09}_{-0.09}$
	 & $1.60^{+0.23}_{-0.21}$ \\
\hline\hline
\multicolumn{6}{c}{Two temperature fitting} \\ \hline
field & O/Fe & Ne/Fe & Mg/Fe & Si/Fe & S/Fe \\ \hline
center   & $0.52^{+0.13}_{-0.12}$	 & $1.85^{+0.33}_{-0.30}$
	 & $1.20^{+0.13}_{-0.12}$	 & $0.71^{+0.07}_{-0.07}$
	 & $0.85^{+0.10}_{-0.09}$ \\
offset   & $0.61^{+0.40}_{-0.17}$	 & $2.22^{+0.85}_{-0.78}$
	 & $1.42^{+0.40}_{-0.30}$	 & $0.61^{+0.18}_{-0.15}$
	 & $1.13^{+0.43}_{-0.35}$ \\
all      & $0.56^{+0.21}_{-0.10}$	 & $2.03^{+0.46}_{-0.42}$
	 & $1.31^{+0.21}_{-0.16}$	 & $0.66^{+0.10}_{-0.08}$
	 & $0.99^{+0.22}_{-0.18}$ \\
\hline
\end{longtable}

\subsection{Uncertainties in the spectral Fits}
\label{sec:res_sys_err}

To estimate the systematic errors, we varied the thickness of the OBF
contaminant and the NXB levels by $\pm 10\%$ in the same manner as in
\citet{kSato2008b}.  The results are summarized in Figure
\ref{fig:kt_sys} (a) and Figure \ref{fig:abund_sys}.
Here, the values in the north and east offset regions ($r > 9'$) are
averaged.  Table \ref{tab:chi_sys} summarizes $\chi^2$ values.
The 10\% decrease in the OBF contaminant gives a slightly better
$\chi^2$ value, while the change in the NXB level has little
effect.

We note that the systematic uncertainty in the derived temperature due
to the change in the OBF contaminant is less than the statistical
error in all the observed regions.  In particular, the effect on the
cool ICM component is almost negligible.  The 10\% change in the NXB
level gives nearly the same ICM temperature and the effect is
insignificant.

The change in the metal abundance due to the 10\% change in the OBF
contaminant is smaller than 10--20\%
 for all the elements, and it
is very close to the statistical error.
The change in the NXB level again has a negligible effect on the
abundance determination.

In summary, the uncertainty in the NXB level and the OBF contaminant
have little effect on the temperature and abundance structures.

\begin{longtable}{lccccc}
\caption{List of $\chi^2/$dof for each fit to study systematic uncertainties}
\label{tab:chi_sys}
  \hline              
\endfirsthead
  \hline
\endlastfoot
\multicolumn{6}{c}{Single temperature fitting} \\ \hline
Region & Nominal &
 \multicolumn{2}{c}{Contaminant} &
 \multicolumn{2}{c}{NXB} \\
(arcmin/$r_{180}$)
 & - & $+10\%$ & $-10\%$ & $+10\%$ & $-10\%$ \\ \hline
$0 - 2 / 0 - 0.025$
    & $1574/899$ & $1685/899$ & $1483/899$ & $1577/899$ & $1571/899$ \\
$2 - 4 / 0.025 - 0.050$
    & $1441/899$ & $1514/899$ & $1389/899$ & $1447/899$ & $1436/899$ \\
$4 - 6 / 0.050 - 0.076$
    & $1194/899$ & $1239/899$ & $1160/899$ & $1203/899$ & $1187/899$ \\
$6 - 9 / 0.076 - 0.11$
    & $1027/899$ & $1055/899$ & $998/899$  & $1030/899$ & $1019/899$ \\
 \hline
$< 15 / 0.19$ (N)
 & $1047/899$ & $1068/899$ & $1034/899$ & $1070/899$ & $1036/899$ \\
$< 15 / 0.19$ (E)
 & $963/899$  & $999/899$  & $937/899$  & $990/899$  & $949/899$  \\
$> 15 / 0.19$ (N)
 & $1091/899$ & $1095/899$ & $1094/899$ & $1100/899$ & $1095/899$ \\
$> 15 / 0.19$ (E)
 & $969/899$  & $983/899$  & $965/899$  & $994/899$  & $964/899$  \\
\hline\hline
\multicolumn{6}{c}{Two temperature fitting} \\ \hline
Region & Nominal &
 \multicolumn{2}{c}{Contaminant} &
 \multicolumn{2}{c}{NXB} \\
(arcmin/$r_{180}$)
 & - & $+10\%$ & $-10\%$ & $+10\%$ & $-10\%$ \\ \hline
$0 - 2 / 0 - 0.025$
    & $1276/897$ & $1317/897$ & $1245/897$ & $1277/897$ & $1275/897$ \\
$2 - 4 / 0.025 - 0.050$
    & $1206/897$ & $1239/897$ & $1183/897$ & $1209/897$ & $1204/897$ \\
$4 - 6 / 0.050 - 0.076$
    & $1122/897$ & $1148/897$ & $1105/897$ & $1129/897$ & $1118/897$ \\
$6 - 9 / 0.076 - 0.11$
    & $967/897$  & $981/897$  & $957/897$  & $973/897$  & $964/897$  \\
 \hline
$< 15 / 0.19$ (N)
 & $999/897$  & $1001/897$ & $1002/897$ & $1008/897$ & $999/897$  \\
$< 15 / 0.19$ (E)
 & $927/897$  & $946/897$  & $915/897$  & $943/897$  & $921/897$  \\
$> 15 / 0.19$ (N)
 & $1111/897$ & $1099/897$ & $1117/897$ & $1109/897$ & $1115/897$ \\
$> 15 / 0.19$ (E)
 & $967/897$  & $975/897$  & $965/897$  & $986/897$  & $964/897$  \\
\hline
\end{longtable}

\subsection{Gas mass profile derived from the XMM-Newton observation}
\label{sec:res_gas_mass}
To derive an accurate gas mass profile, 
 we used  MOS and PN data 
from the XMM-Newton observation of NGC~5044.
First, to derive accurate normalization for the hot and cool ICM
components, we fitted the MOS and PN spectra, which are accumulated
within thinner annuli, with the best-fit 2T model derived from the
Suzaku spectral fit (\ref{sec:ana_fit}).
We used the results of the central four annuli of the Suzaku spectra, or
 within 
$\sim 0.11~r_{180}~(100~{\rm kpc})$.
The temperature and abundances of the ICM
components were fixed to the best-fit values derived in Section
\ref{sec:res_abund}, and normalizations of the {\it vapec} components
were left free.
To keep the spectral shape of the Galactic and the ICM emission 
 exactly the same as the Suzaku fit, we fixed the normalization
ratio for the two components in the Galactic emission, as well as for
the two ICM components, in all the annuli.

We then calculated a volume filling factor $f(r)$ of the cool
component \citep{Morita2006}. The fractional volumes of the
cool and hot gas, $V_{\rm cool}$ and $V_{\rm hot}$, to the total
volume $V$ are given by $V_{\rm cool} = f V$ and $V_{\rm hot} = (1-f)
V$, respectively.  Because the normalizations of the 2T {\it vapec}
models are expressed as $Norm_1 = C_{12}~n_1^2~V_1$ and $Norm_2 =
C_{12}~n_2^2~V_2$ using a certain common constant $C_{12}$, the volume
filling factor $f(r)$ can be solved under the pressure balance as
\citep{Morita2006},
\begin{eqnarray}
  f(r) = \left[ 1+ \left( \frac{T_{\rm hot}}{T_{\rm cool}} \right) ^2
	\frac{Norm_{\rm hot}}{Norm_{\rm cool}} \right]^{-1}.
\end{eqnarray}

\begin{figure*}
  \begin{center}
    \FigureFile(80mm,60mm){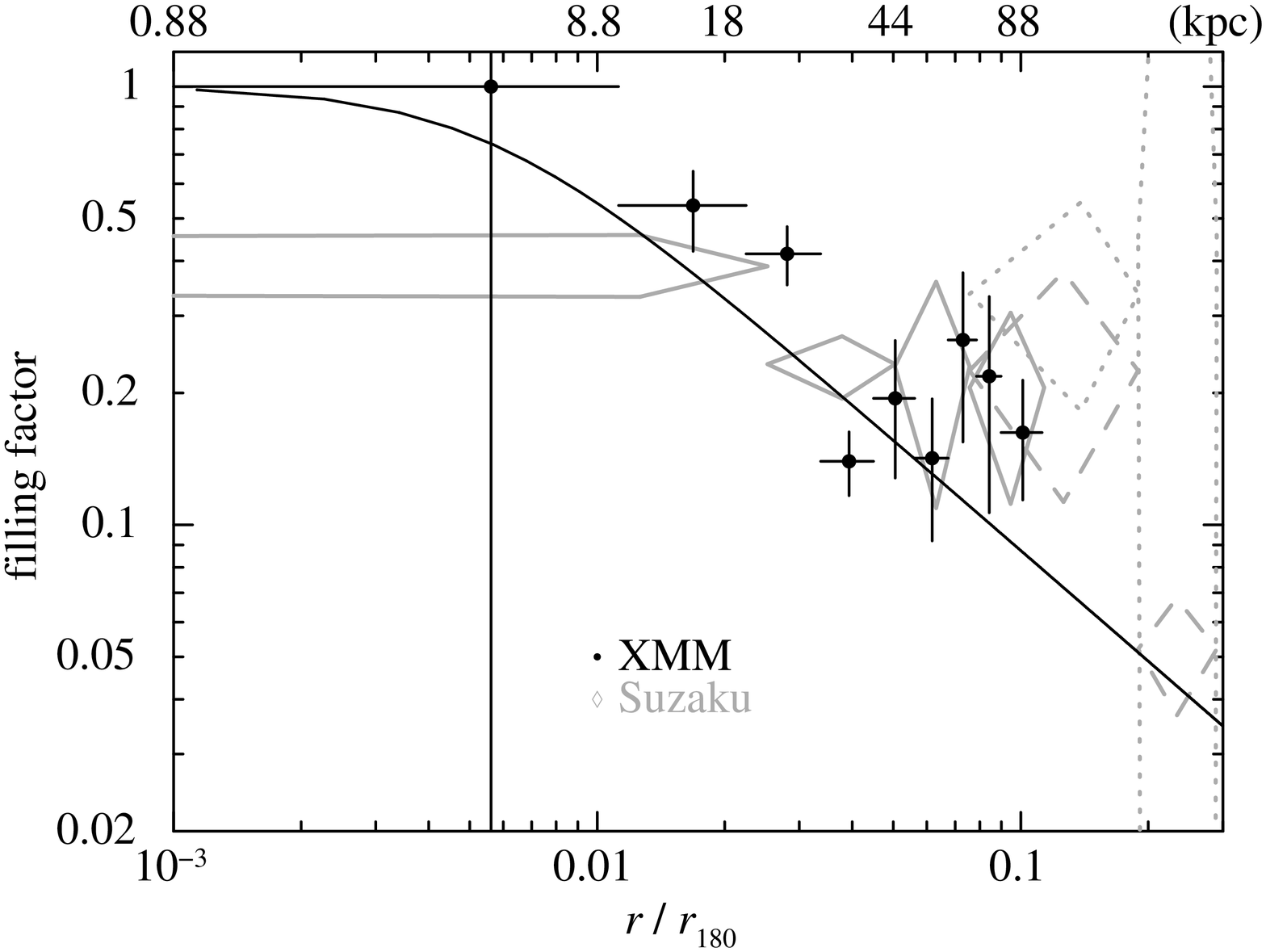}
    \FigureFile(80mm,60mm){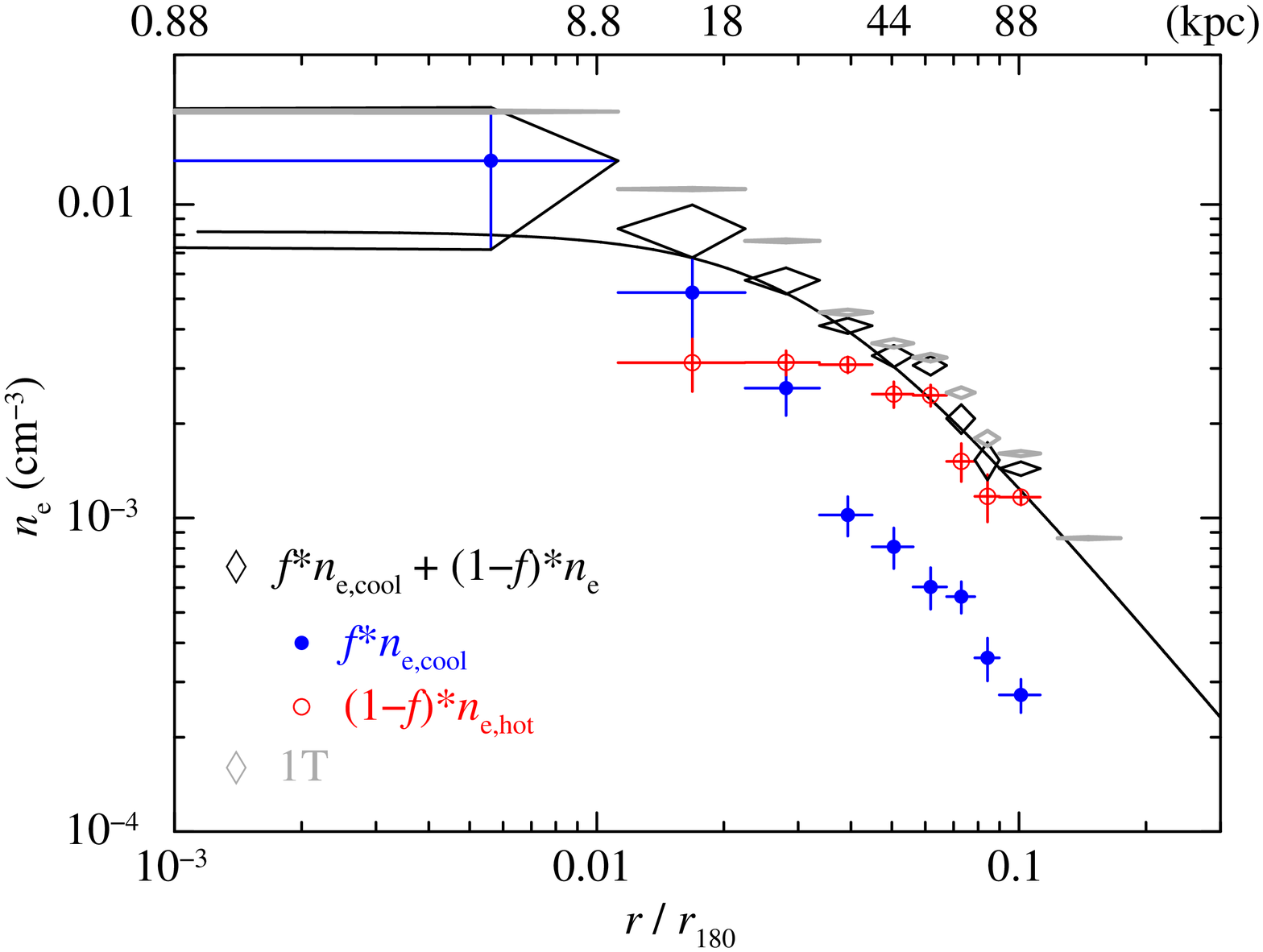}
  \end{center}
  \caption{ {\bf (Left)} Filling factor of the cool component
derived from XMM (black) and Suzaku (gray), plotted along
with the best-fit function
of the results from XMM (black solid-line).  
The Suzaku data for the central four annuli,
 the north field, and east field are shown in
 solid, dashed and dotted lines, respectively. 
{\bf (Right)} Electron
density $n_{\rm e}~({\rm cm}^{-3})$ obtained using the results
 of the 1T and 2T fitting.  
The results for the cool (blue closed circle) and
 hot (red opened circle) ICM components and
 the total for the 1T (gray) and 2T (black)
 models are plotted.  The best-fit $\beta$-model
 of the total 2T model is indicated by
 a solid black line.}
  \label{fig:fil_rho_gas}
\end{figure*}

The radial profile of the filling factor is shown in Figure
\ref{fig:fil_rho_gas} (left), for the XMM and Suzaku data.
Considering the difference in the point spread function between the
XMM and Suzaku data, the two results of the filling factor agree 
well.  
Within 
$r < 0.07~r_{180}$,
 the cool ICM component is more
 dominant than in the outer region. 
This filling factor profile has
 been fitted with a $\beta$-model function,
 $f(r) = \left[ 1+ (r / r_{{\rm c,}f})^2 \right]^{-3\beta_{f}/2}$,
 and we obtained
$r_{{\rm c,}f} = 0.0055~r_{180}~(4.85~{\rm kpc})$
 and $\beta_{f} = 0.28$.

Based on the filling factor result, we derived the gas density
distribution in the same way as in \citet{Morita2006} assuming that two
temperature components of the ICM exist in 
the inner part of the extracted regions, or
$r < \sim 0.11~r_{180}$. 
In the outer region $r > \sim 0.11~r_{180}$,
it is not clear whether  two ICM components exist.
The derived radial profile of electron density
$n_{\rm e}$ is shown in Figure \ref{fig:fil_rho_gas} (right).  
The best-fit model for the filling factor
 is significantly smaller than the observed data within
 0.05-0.2$r_{180}$ (Figure \ref{fig:fil_rho_gas}).
However, this discrepancy does not affect the 
the total electron density within several \%, since
in this region,  the electron density
of the hot component dominates.
Then, we
fitted this density profile with a $\beta$-model function,
 $\rho_{\rm gas}(r) = \rho_{\rm gas,0} \left[ 1+ (r / r_{\rm c})^2
\right]^{-3\beta/2}$, and obtained
 $\rho_{\rm gas,0} = 2.4 \times 10^5~(M{\rm _{\solar}/kpc^3)}$,~
$r_{\rm c} = 0.032~r_{180}~(28~{\rm kpc})$
 and $\beta = 0.53$.

The energy spectra for the XMM-Newton observation were also fitted
with the same model as in our 1T fitting (Sec. \ref{sec:ana_fit}).  As
mentioned above, the single-temperature ICM component was assumed to
exist in all extracted regions, or
$r < \sim 0.17~r_{180}$. 
The temperature and abundances of the ICM component were fixed to the
best-fit values derived in Section \ref{sec:res_abund}, and
normalization of the {\it vapec} component was left free.  We fitted
the gas density profile, that was calculated from the normalizations
derived from the 1T spectral fit, with a $\beta$-model function by
fixing $\beta = 0.51$, which was obtained from the fitting of the XMM
surface brightness profile. The derived parameters of the $\beta$-model
are $\rho_{\rm gas,0} = 6.3 \times 10^5~(M{\rm _{\solar}/kpc^3)}$ and
$r_{\rm c} = 0.017~r_{180}~(15~{\rm kpc})$.

\begin{figure*}
  \begin{center}
    \FigureFile(80mm,60mm){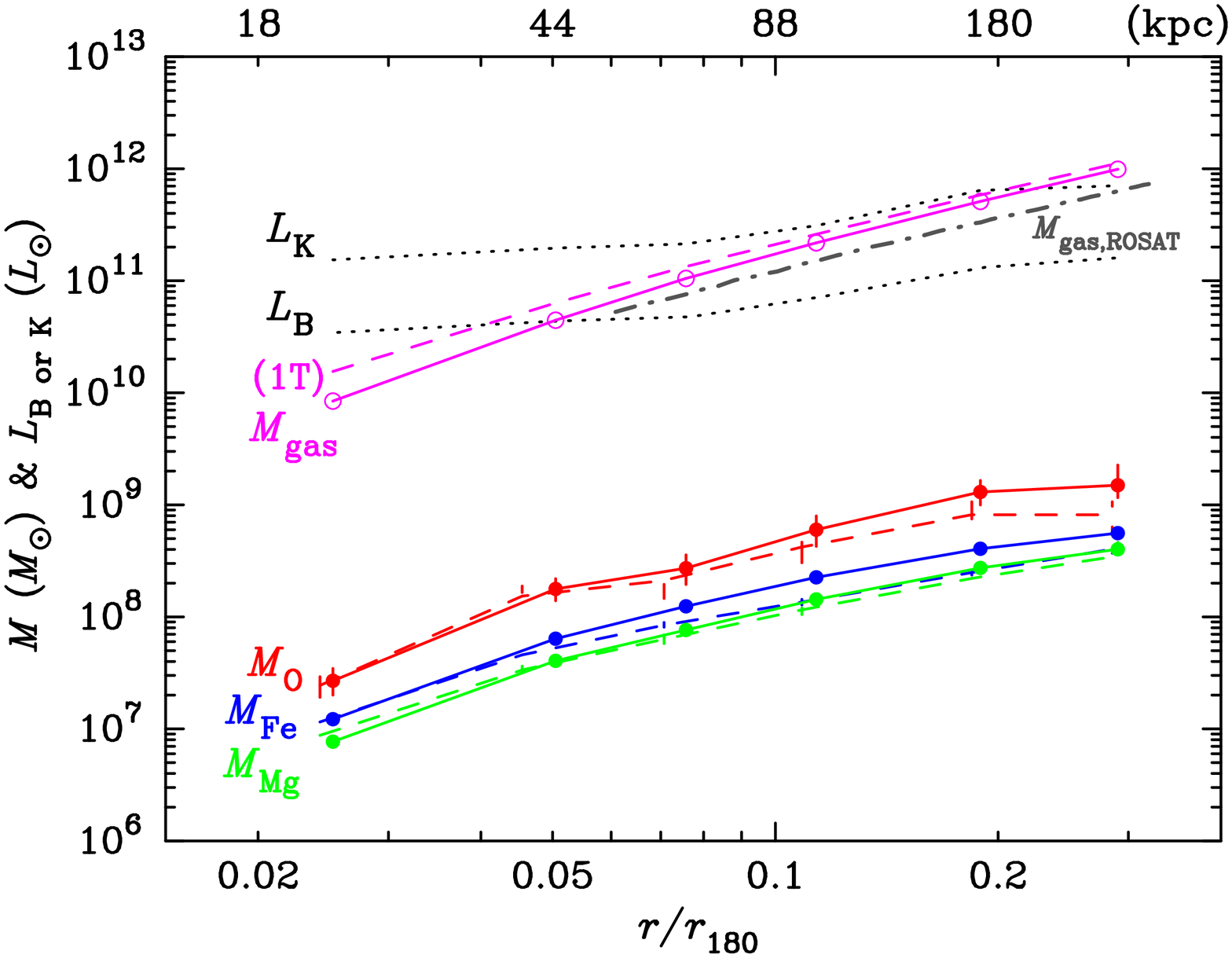}
    \FigureFile(80mm,60mm){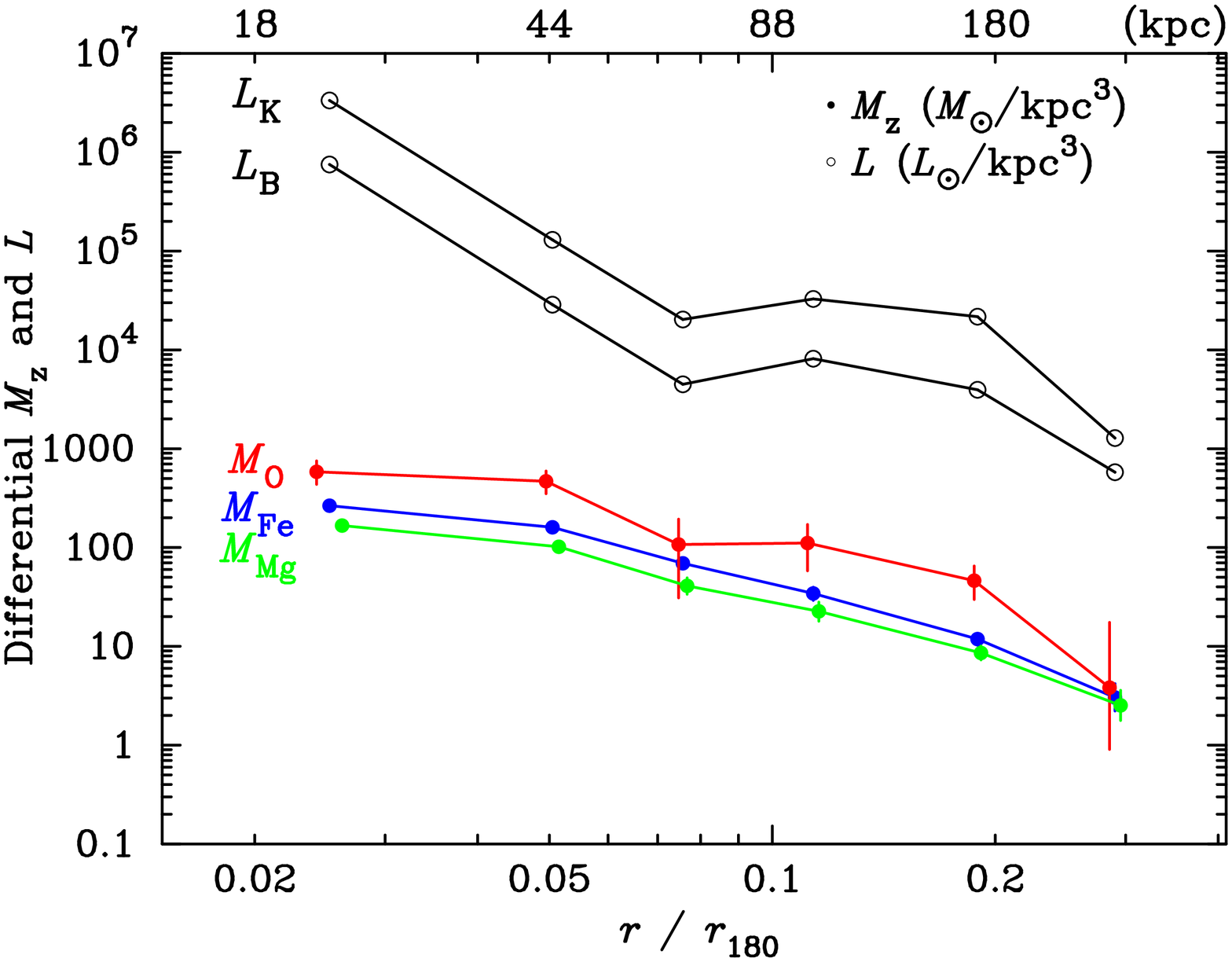}
  \end{center}
  \caption{({\bf Left}) Integrated mass of the hot gas (magenta), 
and those of O (red), Fe (blue), and Mg (green).
Integrated B-band and K-band luminosities of galaxies, $L_{\rm B}$ and
 $L_{\rm K}$ (dotted lines), are also plotted. 
The dashed and solid lines correspond to the mass derived
from the 1T and 2T model fit, respectively.
The  gray dot-dashed line shows the gas mass
derived using ROSAT (\cite{David1994}).
({\bf Right}) Differential metal mass density profiles (blue, red and green
 solid lines) derived for the 2T 
model and  luminosity density profiles of  $L_{\rm B}$, and $L_{\rm K}$. }
  \label{fig:gas_mass}
\end{figure*}

By integrating the gas density $\rho_{\rm gas}$, we derived the
integrated gas mass profile for the 1T and 2T models (Figure
\ref{fig:gas_mass}). For the 2T model, the sum of the cool and hot
components is plotted.  At 
$r > 0.11~r_{180}$ for the 2T fit and 
$r > 0.17~r_{180}$ for the 1T fit,
 we extrapolated the gas mass profile
using the best-fit $\beta$ model.  A clear difference in the gas mass
profile between the 1T and 2T models is seen in the central region,
$r <~\sim 0.03~r_{180}$, while at $r>~\sim 0.03~r_{180}$, the two
models give similar gas mass profiles.  Within $2'$ from the center,
the temperatures of the ICM derived from the 1T model and from the
cool component of the 2T model, which contributes  most of the Fe-L
emission, are nearly the same.  In contrast, the 1T model gives a
lower Fe abundance than the 2T case.  To reproduce 
similar intensity in the Fe-L band, a higher gas mass is required for
the 1T model.

The gas mass profile derived from ROSAT observation
\citep{David1994} is also shown in Figure \ref{fig:gas_mass}.  The
radial gradient of our gas mass profile agrees well with the ROSAT
result, while the absolute values are larger than the ROSAT level by
several tens of percentages, which is probably due to  different assumptions about
 the metal abundance.

Integrated and differential mass profiles of O, Mg, and Fe were derived
from the gas mass and  abundance profiles (Figure
\ref{fig:gas_mass}).  In contrast to the gas mass profile, the metal
mass profiles derived from the 1T and 2T models agree well in the
central region.
Hereafter, we use the metal mass profiles derived from the 2T model.

\section{Discussion}
\label{sec:dis}

\subsection{The Fe abundance profiles of the ICM}

\begin{figure}
  \begin{center}
    \FigureFile(80mm,60mm){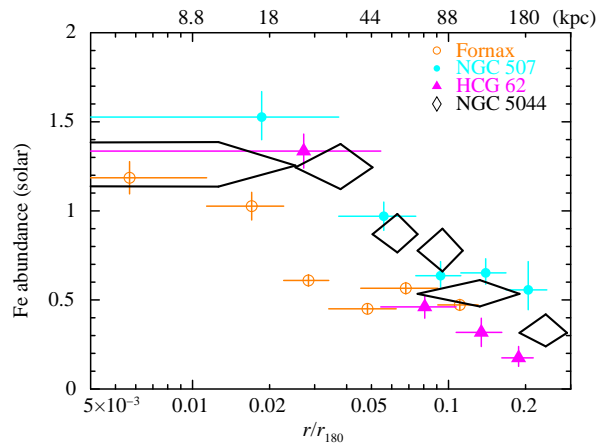}
  \end{center}
  \caption{Radial profiles of the Fe abundance in the NGC 5044 group (diamonds),
the Fornax cluster (\cite{Matsushita2007a}, open circles), the NGC 507 group
 (\cite{kSato2008b}, closed circles), and HCG 62(\cite{Tokoi2008}, closed triangles).}
  \label{fig:fe}
\end{figure}

The radial profile of  Fe abundance in the NGC 5044 group is
compared with those in other groups, HCG 62 (\cite{Tokoi2008}), the NGC
507 group (\cite{kSato2008b}), and the Fornax cluster
(\cite{Matsushita2007a}) observed with Suzaku (Figure \ref{fig:fe}).
The three groups show similar Fe abundance profiles.  The Fe abundance
is 1.2--1.5 solar at the center, and gradually decreases to 0.5 solar
at $0.1r_{180}$.  The value of 0.5 solar is close to the level observed in
general clusters of galaxies at 0.1--0.3$r_{180}$
(\cite{Matsushita2008}).  These groups are characterized by nearly
symmetric X-ray morphology with a bright central galaxy, and the
enhanced Fe abundance within 0.1 $r_{180}$ should reflect  Fe
accumulation in the central galaxies over a long time.  The NGC 5044
group shows similar Fe abundance in the north and east fields,
suggesting that the group gas has evolved by keeping global spatial
symmetry.  In contrast, the Fornax cluster shows nearly constant Fe
abundance at 0.5 solar over a radial range of 0.02--0.13 $r_{180}$
from the cD galaxy, NGC 1399.  This cluster has asymmetric X-ray
morphology, and NGC 1399 is significantly offset from the center
(\cite{Paolillo2002}, \cite{Sharf2005}). Chandra observations
suggest that NGC 1399 may be relatively moving against the ICM
(\cite{Sharf2005}).  Therefore, around NGC 1399, a recent supply of
metals from the cD galaxy produces a sharp peak of Fe abundance within
0.02 $r_{180}$.

\subsection{The abundance pattern of the ICM and contributions from SN
  Ia and SN II}
\label{sec:dis_abund}

\begin{figure}
  \begin{center}
    \FigureFile(80mm,60mm){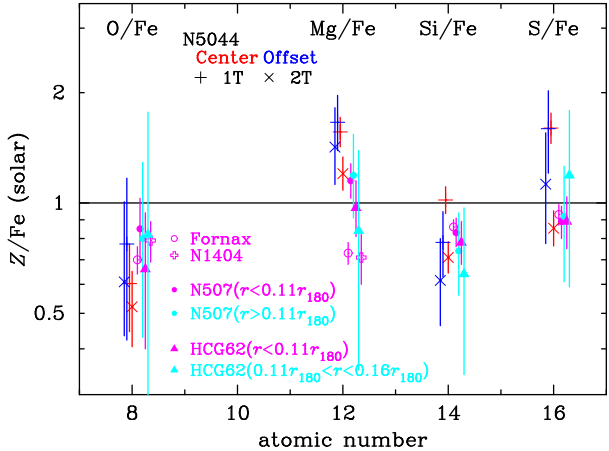}
  \end{center}
  \caption{Weighted averages of the abundance ratios of various elemental species compared to Fe
for the central (red; $r<\sim0.1r_{180}$) and offset (blue; $r>\sim
 0.1r_{180}$) fields of the NGC 5044 group. Those in the regions within
 $\sim 0.1r_{180}$ (magenta) and $0.1\sim 0.3r_{180}$ (light blue)  of the NGC 507 group
(\cite{kSato2008b}), HCG 62 (\cite{Tokoi2008}), the Fornax cluster
 (\cite{Matsushita2007a}), and an elliptical galaxy NGC 1404
 (\cite{Matsushita2007a})
are also plotted.}
  \label{fig:atomnum}
\end{figure}

Figure \ref{fig:atomnum} summarizes the abundance pattern for O, Mg,
Si, S and Fe in four systems: the NGC 5044 group, HCG 62, the 
NGC 507 group
and the Fornax cluster.  The derived ratios, O/Fe, Mg/Fe, Si/Fe and
S/Fe, all scatter around unity, i.e.,\ the solar ratio.  The four systems
show similar Si/Fe and S/Fe ratios.  The NGC 5044 group shows the smallest
O/Fe ratio, 0.6, compared with $\sim 0.8$ in the other three systems.  In
contrast, the NGC 5044 group shows the highest Mg/Fe ratio of 1.2 among
the four systems derived from the 2T model.  The 2T and 1T models give
different abundance ratios by a few tens of percentages.  The uncertainty due to
coupling with the Fe-L lines may result in a 20--30\% systematic error in
the Mg abundance \citep{Matsushita2007a}, and the estimation of the
Galactic emission leaves some uncertainty in the O
abundance. 
Considering the systematic uncertainty  mentioned here,
the abundance pattern for O, Mg, Si, S, and Fe does not differ
significantly among the four poor systems.

\begin{figure*}
  \begin{center}
  \FigureFile(80mm,80mm){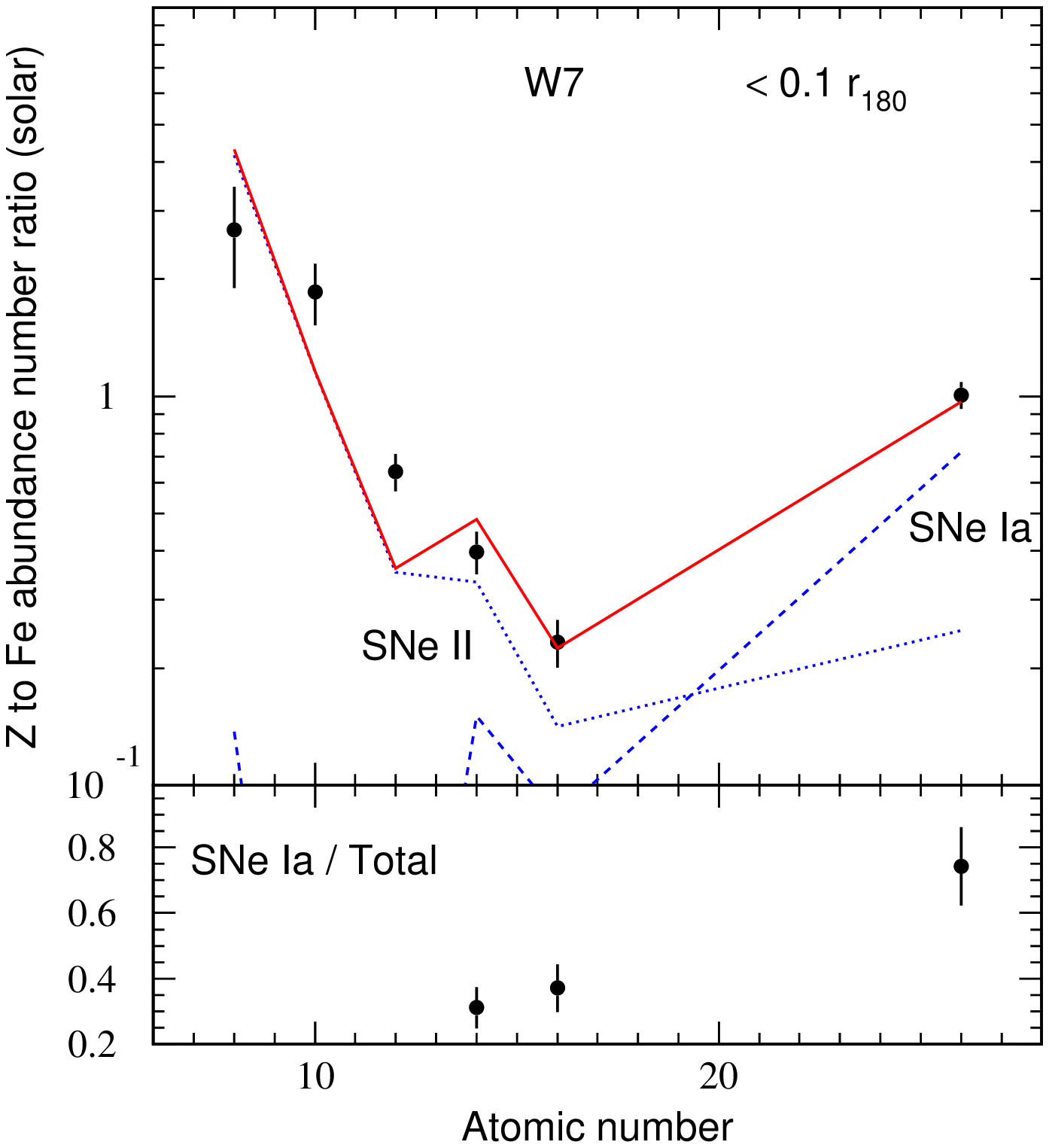}
  \FigureFile(80mm,80mm){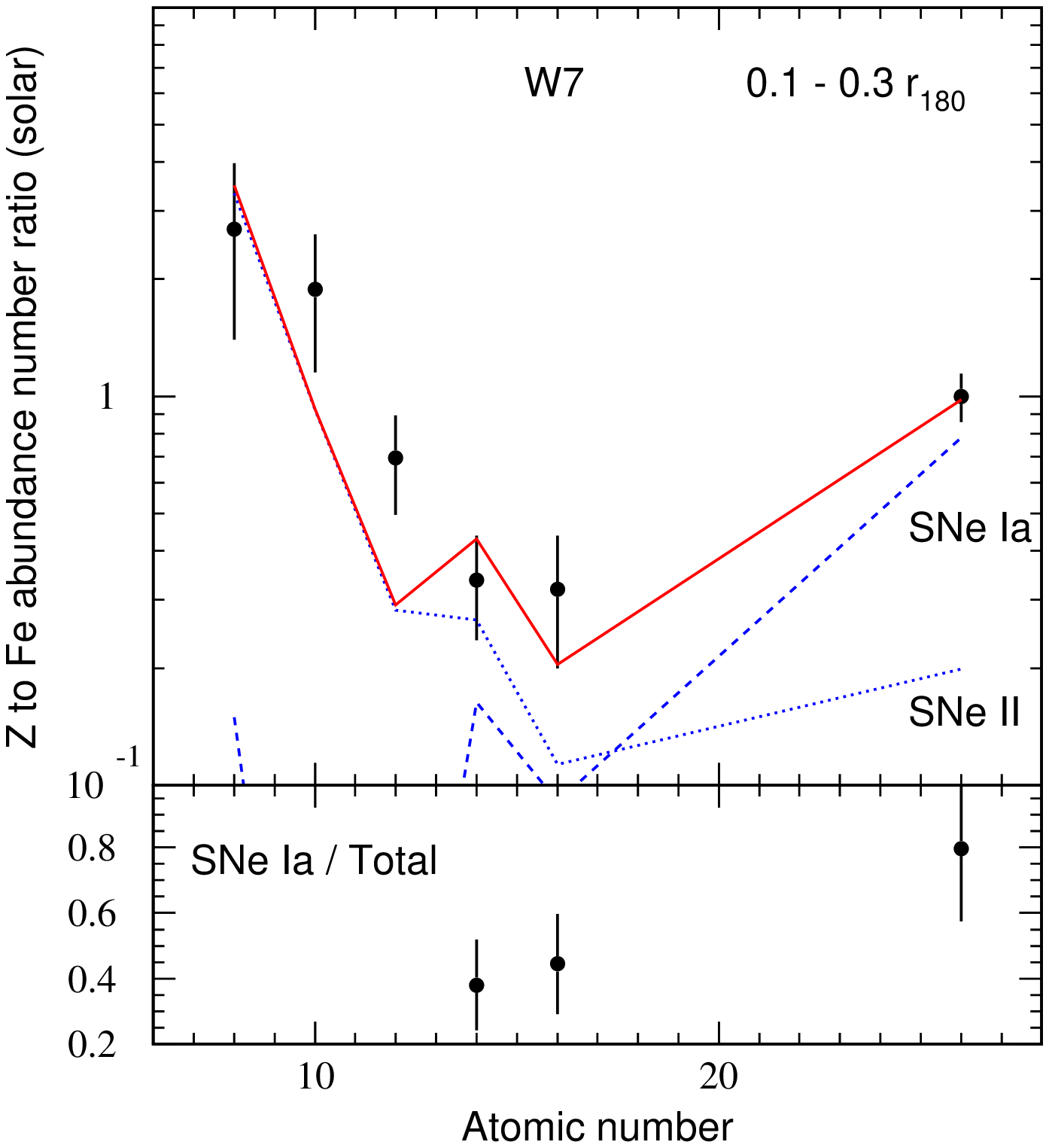}
  \end{center}
  \caption{ Fit results of number ratios of
 elements to Fe of the NGC 5044 group within 0.1 $r_{180}$ {\bf (left)}
and 0.1--0.3 $r_{180}$ {\bf (right)}.
 Top panel shows the number ratios (closed circles).
  Blue dashed and dotted lines correspond to the contributions of SNe Ia
 (W7) and SNe II, respectively, and the 
red line corresponds to the sum of
 the two.
 Ne (atomic number = 10) is excluded from the fit. 
The bottom panel
 indicates fractions of the SNe Ia contribution to
 total mass  in the best-fit model for each element. }
  \label{fig:atomfit}
\end{figure*}

As shown by \citet{kSato2007b}, the abundance pattern of the NGC 5044 
ICM within 0.1$r_{180}$ and 0.1--0.3$r_{180}$ can be fit
 by a combination of average SNe Ia
and SNe II yields (Figure \ref{fig:atomfit}).  The SNe Ia and II
yields were taken from Iwamoto et al. (1999) and Nomoto et al. (2006),
respectively.  About  $80\%$ of Fe and $\sim 40\%$
of Si and S in the ICM are synthesized by SN Ia.  The derived number
ratio of SNe II to SNe Ia is $ 3.1\pm0.6$ and 3.3$\pm$0.7,
for $r<0.1r_{180}$  and $0.1r_{180}<r<0.3r_{180}$, respectively,
assuming a classical
deflagration model, W7, for SNe Ia.  This is close to the value of
3--4, obtained for the other groups of galaxies,  NGC 507 and HCG 62, and
 several other clusters of galaxies (\cite{kSato2007b}).

When systems have a similar initial mass function of stars (IMF) and
are old enough,  most  SN Ia and SN II would have already
exploded.  
To explain  abundance pattern of stars in the solar
neighborhood, lifetimes of SN Ia are confined within 0.5--3 Gyr,
with typical lifetime of 1.5 Gyr
(\cite{Yoshii1996}).
In clusters of galaxies,  to account for Fe mass in the ICM,
the past average rate of SNe Ia was 
 much larger than the present rate in elliptical galaxies
(e.g., \cite{Renzini1993}).
This result indicates that lifetimes of most of  SN Ia 
are much shorter than the Hubble time.
In this case, the final abundance pattern should be
similar.  If the IMF is close to that in our Galaxy and
most of stars in our Galaxy and clusters were already formed before
a few Gyrs ago, the abundance
pattern should naturally be similar to the solar abundance pattern.
Within $0.1 r_{180}$ of NGC 5044, the gas mass and  stellar mass
are pretty close (\S\ref{sec:dis_mlr}).  The mean stellar metallicity
and [$\alpha$/Fe] derived from optical observations are about 1 solar
and $0.34\pm 0.17$, respectively (\cite{kobayashi1999},
\cite{Ann2007}).  Therefore, the sum of the low O/Fe ratio found in
the ICM and the high stellar $\alpha$/Fe values may give roughly the
solar abundance pattern. However, in this study, we are combining  the abundances
of two distinct bodies, stars and ISM, which may have very different
enrichment histories.

\subsection{The radial profiles of the metal mass-to-light ratios}
\label{sec:dis_mlr}
Since metals in the ICM are all synthesized in galaxies, the
metal-mass-to-light ratio is a useful measure for the study of ICM
chemical evolution.  To estimate the metal mass-to-light ratio, we
calculated B-band and K-band luminosity profiles.
\citet{Ferguson1990} (hereafter, FS90) have optically studied the
NGC~5044 group, and identified 162 member galaxies whose apparent
magnitudes, $m_{\rm B}$, are brighter than $m_{\rm B}\sim20$, with a
completeness limit of $m_{\rm B} \sim 18$.  They also show that the
central elliptical galaxy NGC~5044 has an apparent magnitude of
$m_{\rm B} = 11.9$, or $\log{L_{\rm B}/L_{\rm B,\solar}} = 10.7$ using
the luminosity distance
 $D_{\rm L} = 38.9$ 
 Mpc and the foreground
Galactic extinction $A_{\rm B} = 0.300$ \citep{Schlegel1998} from
NED\@.  Since this galaxy dominates the central region, we applied the
de Vaucouleurs' law to calculate the luminosity profile within 
$0.076~r_{180}~(6')$,
 and deprojected the profile using the expression by
 \citet{Mellier1987}.  
In the outer region, 
$r > 0.076~r_{180}$,
 we integrated the luminosities of the member galaxies from FS90
 and deprojected the profile.

Since the K-band luminosity of a galaxy correlates well with the
stellar mass, we also collected K-band magnitudes of galaxies in a box
of $2 \times 2~{\rm deg}^2$ centered on NGC~5044 from the Two Micron
All Sky Survey (2MASS)\@.  NGC~5044 itself has an apparent magnitude
$m_{\rm K} = 7.845$, or $\log{L_{\rm K}/L_{\rm K,\solar}} = 11.4$
using the foreground Galactic extinction $A_{\rm K} = 0.026$
\citep{Schlegel1998} from NED\@.  The average surface brightness in the
region of $r > r_{180} = 79.2'$, $L_{\rm K,bgd} = 3.01 \times
10^9~L_{\rm K,\solar}$, is subtracted as the background.  We then
calculated the K-band luminosity profile in the same way as for
the B-band profile.

\begin{figure*}
  \begin{center}
    \FigureFile(80mm,80mm){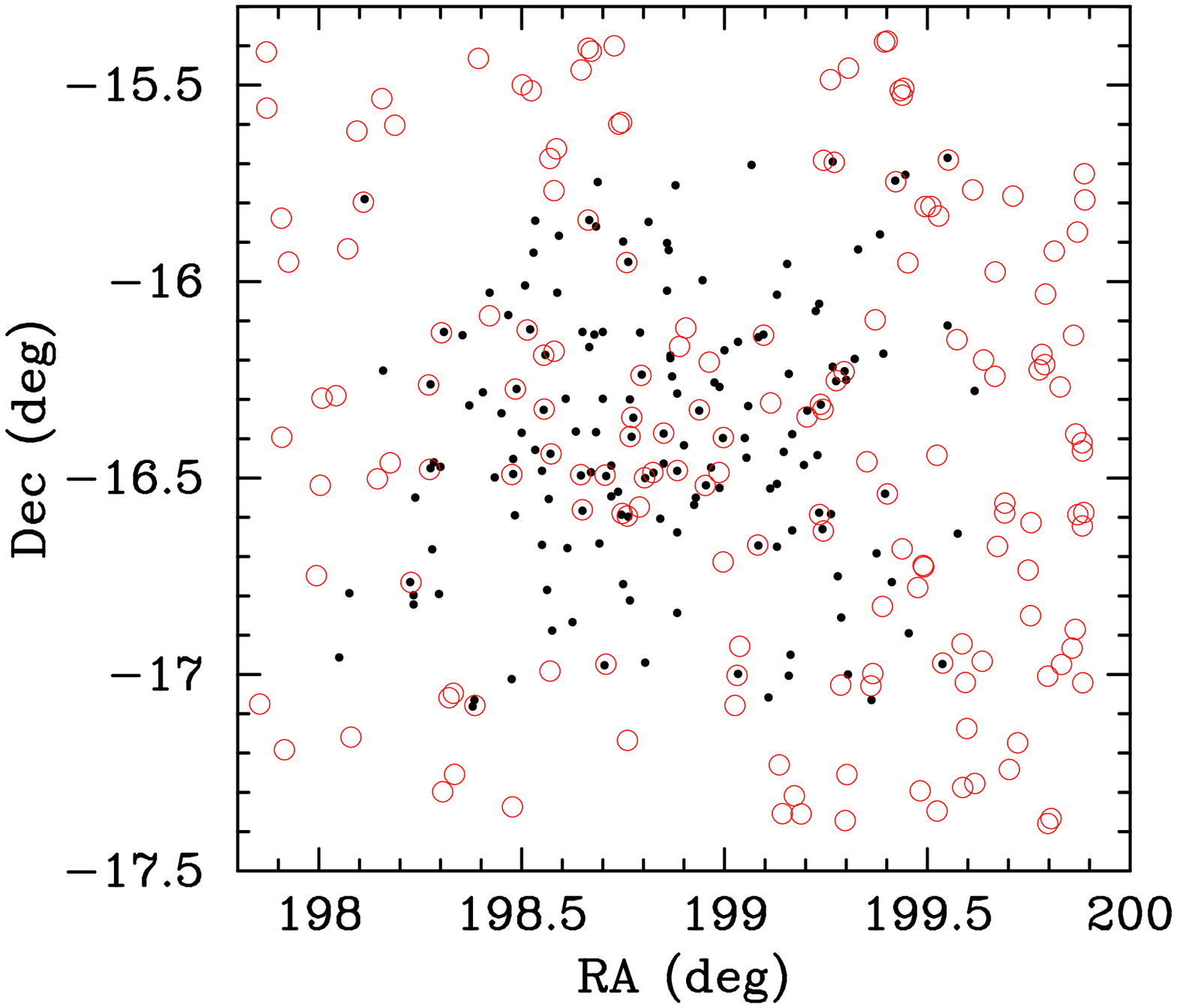}
    \FigureFile(75mm,75mm){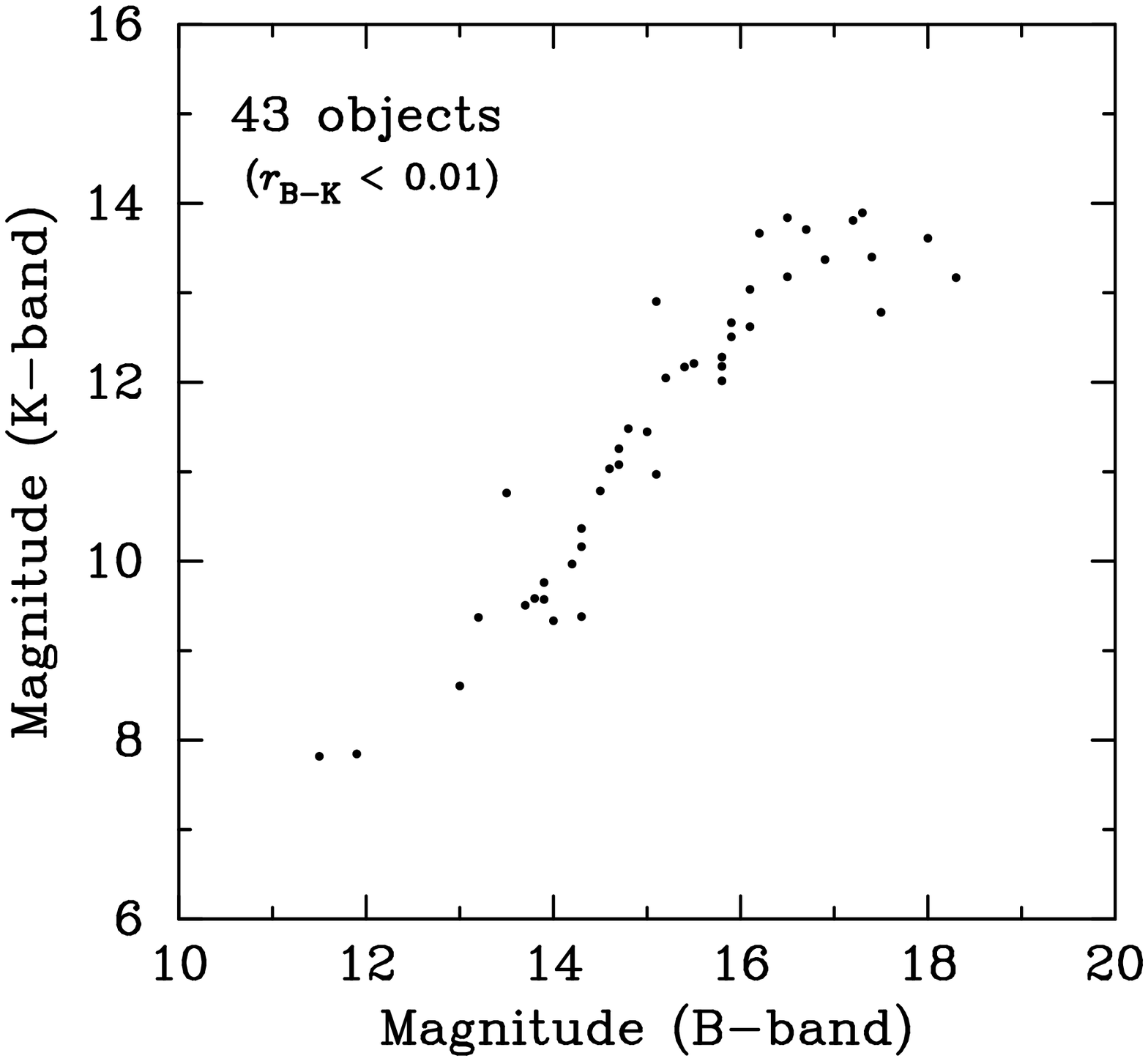}
  \end{center}
  \caption{ {\bf (Left)} Objects detected in B-band and K-band (black
closed circle: B-band,~red open circle: K-band).  {\bf (Right)}
Correlation of magnitude of  B-band and K-band  for  objects detected
in both bands.}
  \label{fig:b-k}
\end{figure*}

Forty-three  objects are detected in both the B-band and K-band
(Figure \ref{fig:b-k}).
As shown in Figure \ref{fig:b-k}, the two magnitudes of these objects
correlate well with each other.

\begin{figure}
  \begin{center}
    \FigureFile(80mm,60mm){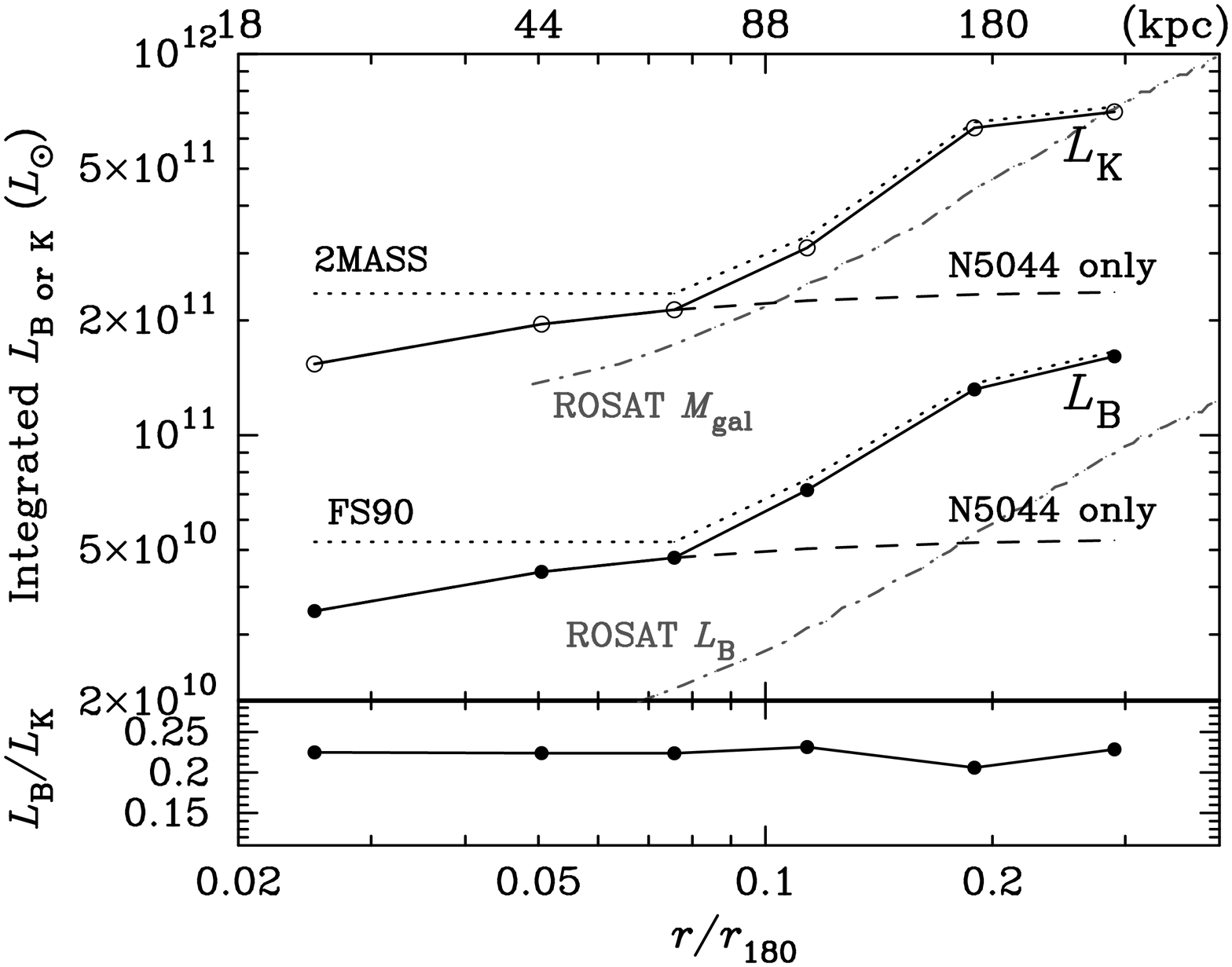}
  \end{center}
  \caption{Integrated luminosity profiles in B-band and K-band.
The dashed lines indicate luminosity 
 of the central galaxy NGC~5044.}
  \label{fig:l_radial}
\end{figure}

The integrated B-band and K-band luminosity profiles are shown in the
left panel of Figure \ref{fig:gas_mass} and in Figure
\ref{fig:l_radial}.
The two band profiles resemble  each other; the ratio of the
integrated luminosity, $L_{\rm B}/L_{\rm K}$, is nearly constant at
$\sim 0.22$ (figure \ref{fig:l_radial}).  This corresponds to $B-K
\sim 4.1$, which is consistent with $B-K = 4.2$ for early-type
galaxies in \citet{Lin2004}.
The NGC~5044 galaxy dominates the luminosity profile in the region $r
< \sim 0.07~r_{180}$, while in the outer region, $r > \sim
0.07~r_{180}$, the contribution from fainter galaxies in the group becomes
important.

The luminosity density profiles of $L_{\rm B}$ and $L_{\rm K}$ are
plotted in the right panel of Figure \ref{fig:gas_mass}.  These
profiles are compared with the mass density profiles of O, Mg and Fe
from the 2T model fit. We note that the mass densities of stars and O
both indicate a shoulder-like structure in the region from
$0.06r_{180}$ to $0.2r_{180}$. The profiles of Mg and Fe do not show such
a shoulder and decrease smoothly. 
Therefore, the oxygen distribution seems to closely correlate with that
of small galaxies in the group.

\begin{figure*}
  \begin{center}
    \FigureFile(80mm,60mm){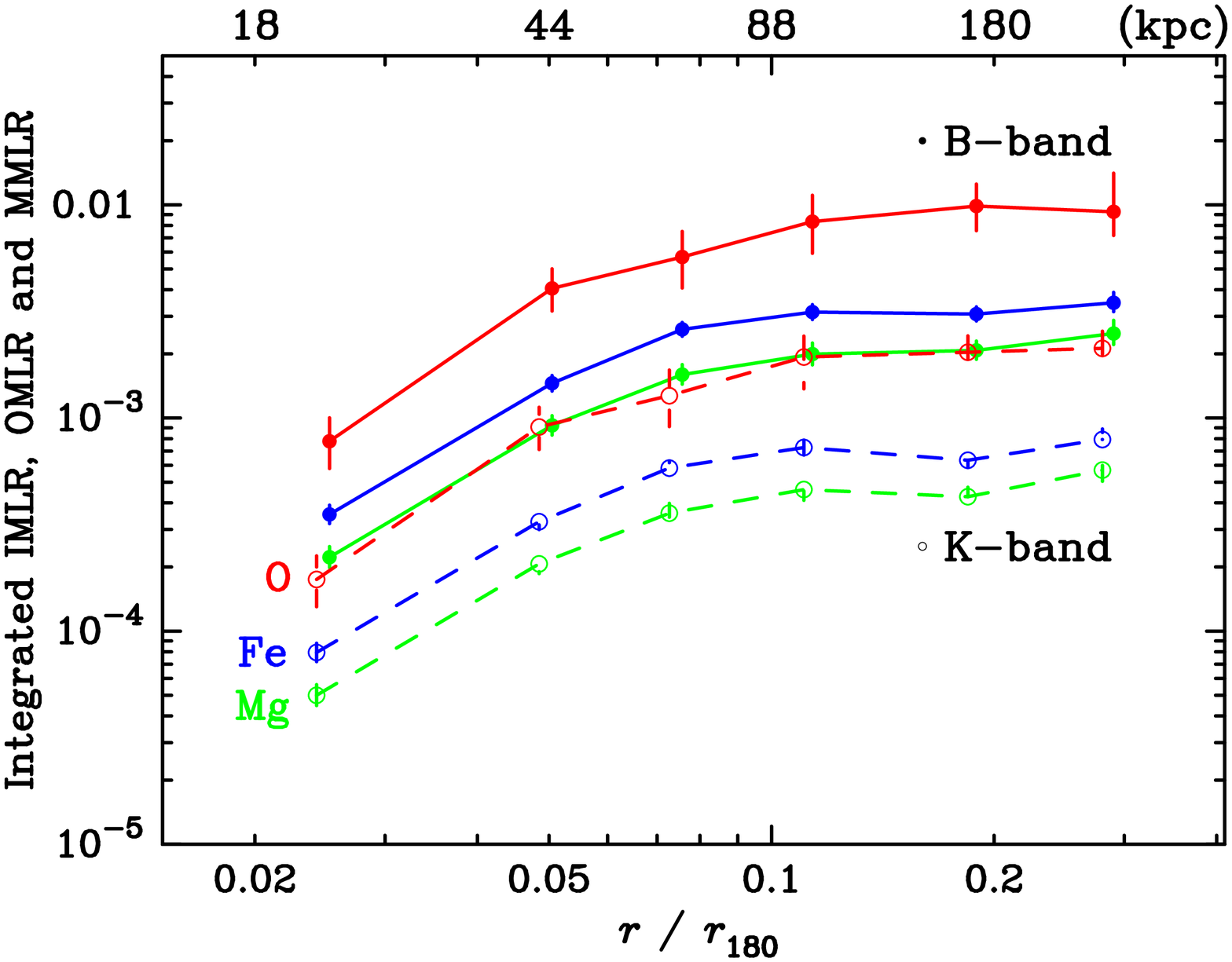}
    \FigureFile(80mm,60mm){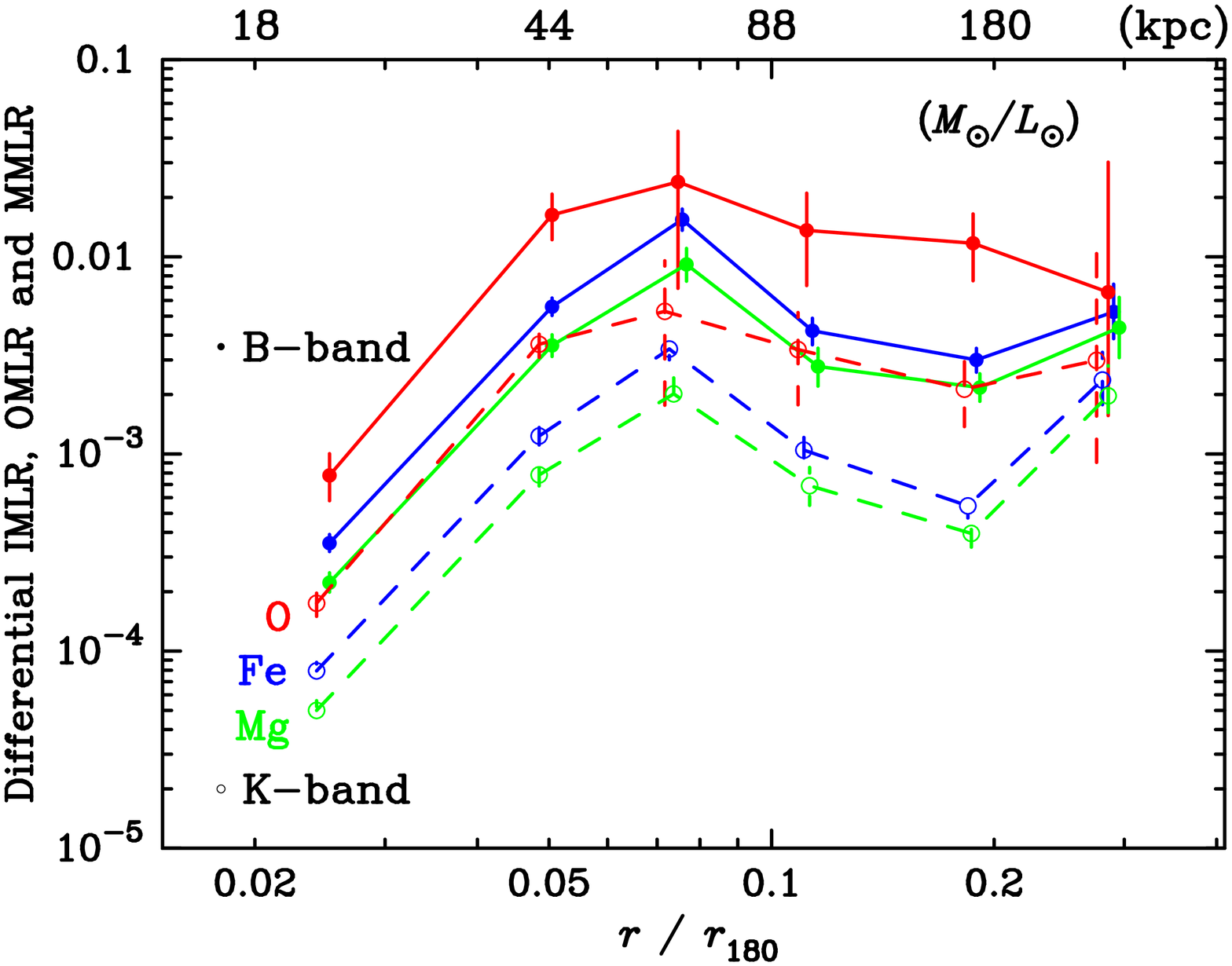}
  \end{center}
  \caption{ {\bf (Left)} Radial profiles of integrated IMLR
(blue), OMLR (red) and MMLR (green) in B-band (solid lines) and K-band
(dashed lines).  {\bf (Right)} Radial profiles of differential
mass-to-light ratios with the same representation as in the left panel.
}
  \label{fig:mlr_radial}
\end{figure*}

The integrated mass-to-light ratios for O, Mg and Fe (OMLR,
 MMLR and IMLR) using B-band and K-band luminosities are summarized in
 Table \ref{tab:mlr_n5044} and Figure \ref{fig:mlr_radial} (left).
The error bars of the mass-to-light ratios include
 only the abundance errors.
The profiles increase with the radius up to $r \sim 0.1~r_{180}$, 
in the region where NGC 5044 dominates, and then become almost
constant in the outer region.

Our IMLR value with the B-band luminosity is $\sim 3.6 \times 10^{-3}$
expressed in terms of  solar values at $r < 0.3~r_{180}$.  
This value is consistent
with the result of \citet{Buote2004}, since our larger derived Fe abundance
gives a smaller gas mass.

We also calculated differential  metal-mass-to-light ratios,
as shown in the right panel of Figure \ref{fig:mlr_radial}.
The differential OMLR becomes almost flat in the region $r > \sim
0.07~r_{180}$, while IMLR and MMLR show a clear drop in
0.1--0.2$r_{180}$ compared with the level at $0.07r_{180}$.
In other words, the O distribution seems to be more extended than those of  
Fe and Mg, although the error in the data is fairly large.  There
may be some difference in the metal production process between the
giant elliptical galaxy, NGC 5044, and smaller galaxies, in connection
with a different history of SN Ia and SN II activity.

\begin{longtable}{cccccccc}
\caption{Summary of metal-mass-to-light ratios of NGC~5044}
\label{tab:mlr_n5044}
  \hline              
\endfirsthead
  \hline
\multicolumn{8}{{@{}l@{}}}{\hbox to 0pt{\parbox{120mm}{\footnotesize
\par\noindent
\footnotemark[$\ast$]
 To compare with the results of other objects
 in Figure \ref{fig:mlr},
 we recalculated metal mass-to-light ratios at $r = 0.1~r_{180}$
 and $r = 0.3~r_{180}$\\
}\hss}}
\endlastfoot
\multicolumn{8}{c}{B-band} \\ \hline
\multicolumn{3}{c}{$r$}
 & $L_{\rm B}$
 & $M_{\rm gas}$
 & IMLR
 & OMLR
 & MMLR
 \\
($r_{180}$) & (arcmin) & (kpc)
 & $10^{10}~L_{\rm \solar}$
 & $10^{11}~M_{\rm \solar}$
 & \multicolumn{3}{c}{$\times 10^{-3}~\left(M_{\rm \solar}/L_{\rm \solar}\right)$} \\
\hline
$0.025$ & $2$ & $22$
 & $3.5$
 & $0.084$
 & $0.35^{+0.04}_{-0.03}$
 & $0.78^{+0.22}_{-0.20}$
 & $0.22^{+0.03}_{-0.02}$
 \\
$0.050$ & $4$ & $44$
 & $4.4$
 & $0.44$
 & $1.5^{+0.1}_{-0.1}$
 & $4.1^{+1.0}_{-0.9}$
 & $0.92^{+0.10}_{-0.09}$
 \\
$0.076$ & $6$ & $66$
 & $4.8$
 & $1.0$
 & $2.6^{+0.2}_{-0.2}$
 & $5.7^{+1.8}_{-1.6}$
 & $1.6^{+0.2}_{-0.2}$
 \\
$0.11$ & $9$ & $100$
 & $7.2$
 & $2.2$
 & $3.1^{+0.3}_{-0.2}$
 & $8.3^{+2.7}_{-2.4}$
 & $2.0^{+0.3}_{-0.2}$
 \\
$0.19$ & $15$ & $167$
 & $13$
 & $5.1$
 & $3.1^{+0.2}_{-0.2}$
 & $9.9^{+2.6}_{-2.3}$
 & $2.1^{+0.2}_{-0.2}$
 \\
$0.29$ & $23$ & $256$
 & $16$
 & $9.9$
 & $3.5^{+0.4}_{-0.3}$
 & $9.3^{+4.8}_{-2.1}$
 & $2.5^{+0.4}_{-0.3}$
 \\ \hline
$0.10$\footnotemark[$\ast$] & $7.9$ & $88$
 & $7.2$
 & $1.7$
 & $2.6^{+0.2}_{-0.2}$
 & $6.6^{+1.9}_{-1.7}$
 & $1.6^{+0.2}_{-0.2}$
 \\
$0.30$\footnotemark[$\ast$] & $24$ & $264$
 & $16$
 & $1.0$
 & $3.6^{+0.4}_{-0.3}$
 & $9.4^{+5.2}_{-2.1}$
 & $2.6^{+0.4}_{-0.3}$
 \\
\hline\hline
\multicolumn{8}{c}{K-band} \\ \hline
\multicolumn{3}{c}{$r$}
 & $L_{\rm K}$
 & $M_{\rm gas}$
 & IMLR
 & OMLR
 & MMLR \\
($r_{180}$) & (arcmin) & (kpc)
 & $10^{11}~L_{\rm \solar}$
 & $10^{11}~M_{\rm \solar}$
 & \multicolumn{3}{c}{$\times 10^{-4}~\left(M_{\rm \solar}/L_{\rm \solar}\right)$} \\
\hline
$0.025$ & $2$ & $22$ 
 & $1.5$
 & $0.084$ 
 & $0.79^{+0.08}_{-0.07}$
 & $1.7^{+0.5}_{-0.4}$
 & $0.50^{+0.06}_{-0.05}$
 \\
$0.050$ & $4$ & $44$
 & $2.0$
 & $0.44$
 & $3.3^{+0.3}_{-0.3}$
 & $9.1^{+2.1}_{-2.0}$
 & $2.1^{+0.2}_{-0.2}$
 \\
$0.076$ & $6$ & $66$
 & $2.1$
 & $1.0$
 & $5.8^{+0.5}_{-0.4}$
 & $13^{+4}_{-4}$
 & $3.6^{+0.4}_{-0.4}$
 \\
$0.11$ & $9$ & $100$
 & $3.1$
 & $2.2$
 & $7.3^{+0.6}_{-0.6}$
 & $19^{+6}_{-6}$
 & $4.6^{+0.6}_{-0.5}$
 \\
$0.19$ & $15$ & $167$
 & $6.4$
 & $5.1$
 & $6.3^{+0.5}_{-0.5}$
 & $20^{+5}_{-5}$
 & $4.3^{+0.5}_{-0.4}$
 \\
$0.29$ & $23$ & $256$
 & $7.0$
 & $9.9$
 & $7.9^{+0.9}_{-0.7}$
 & $21^{+11}_{-5}$
 & $5.7^{+0.9}_{-0.6}$
 \\ \hline
$0.10$\footnotemark[$\ast$] & $7.9$ & $88$
 & $3.1$
 & $1.7$
 & $6.0^{+0.4}_{-0.4}$
 & $15^{+4}_{-4}$
 & $3.8^{+0.4}_{-0.4}$
 \\
$0.30$\footnotemark[$\ast$] & $24$ & $264$
 & $7.0$
 & $1.0$
 & $8.1^{+1.0}_{-0.8}$
 & $21^{+12}_{-5}$
 & $5.9^{+0.9}_{-0.7}$
 \\
\end{longtable}

\subsection{Comparison of the metal-mass-to-light ratios with other systems}

\begin{figure*}
  \begin{center}
    \FigureFile(80mm,120mm){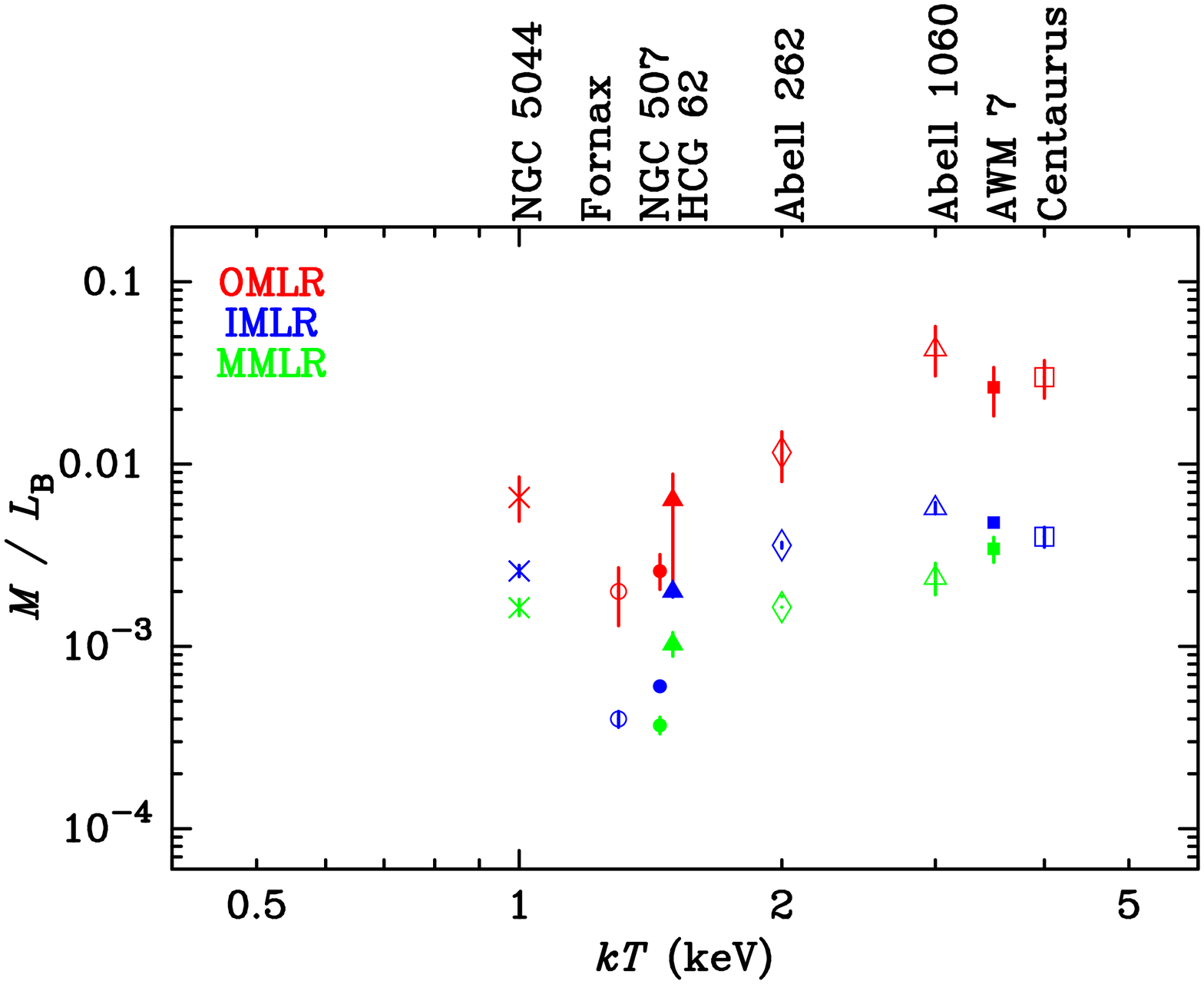}
    \FigureFile(80mm,120mm){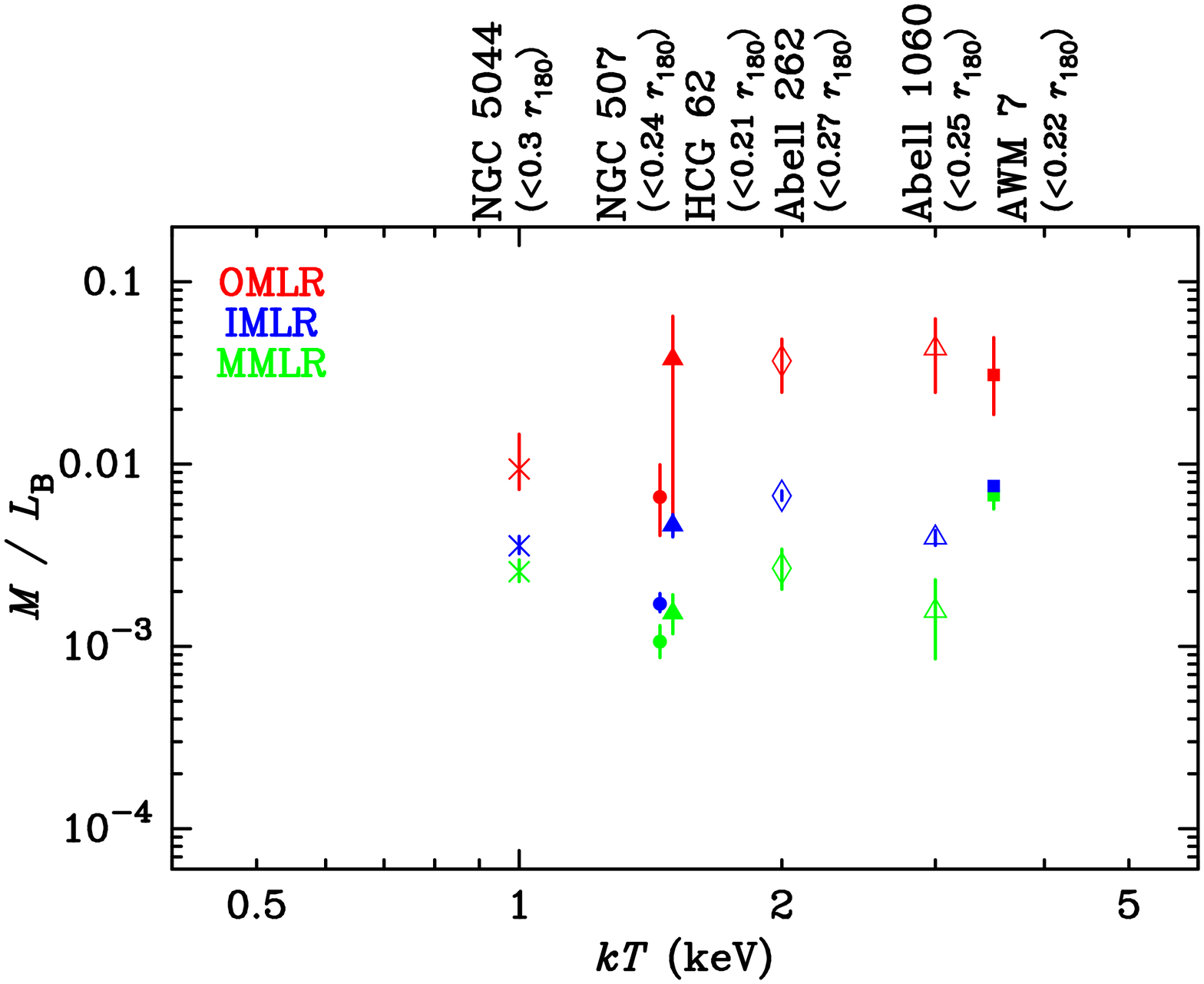}
  \end{center}
  \caption{Integrated IMLR (blue),~OMLR (red) and MMLR (green) at
{\bf (Left)} $r = 0.1~r_{180}$ or 
{\bf (Right)} $r \sim 0.3~r_{180}$
 using the B-band  luminosity.}
  \label{fig:mlr}
\end{figure*}

The metal-mass-to-light ratios of the NGC 5044 group are compared with those in
other groups and clusters. The objects are, the NGC 507 group and HCG 62,
the Fornax cluster as a poor cluster, and other clusters of galaxies
with ICM temperatures of 2--4 keV: A 262, A 1060, AWM7, and the
Centaurus cluster, as summarized in Figure \ref{fig:mlr} and Table
\ref{tab:mlr_all}.
The gas mass of these systems were derived using XMM-Newton data
except the Fornax cluster, where the ROSAT data was used.

The OMLR, MMLR, and IMLR for clusters of galaxies with $kT\sim 2-4$
keV tend to be higher than those in groups of galaxies and the poor
cluster of galaxies with $kT\sim 1$ keV\@. 
However, the data points
might have some systematic uncertainty, possibly because of different
observations of $B$-band luminosity of galaxies, statistics in the
number counts of galaxies, and systematic shift of gas mass derived
from XMM observations with different sensitivities and fields of view.

The difference in the OMLR values between groups and clusters is about
a factor of 3--6, and systematically larger than the IMLR and MMLR
difference, which is a factor of 2--3.  
Among groups, the NGC 5044 group and HCG 62 have the highest IMLRs,
which are comparable to those of clusters.
In contrast, OMLR of the NGC 5044 appears to be lower than those of clusters.
Among the groups and poor
clusters, the Fornax cluster shows the lowest IMLR and OMLR\@.  In the
Fornax cluster, the asymmetric X-ray emission may point to  
large-scale dynamical evolution, which might have hampered the strong
concentration of hot gas in the center.

These poor systems also differ from richer systems in that the gas
density profile in the central region is significantly flatter and
 the relative entropy level is correspondingly higher (e.g.\
Ponman et al.\ 1999).
These peculiar central characteristics in the poor systems are thought
to be best attributed to an injection of energy (preheating) into the
gas before clusters have collapsed (e.g.\ Kaiser 1991). Note also that
more recent observations and simulations of entropy profiles indicate
some modification to this simple picture (e.g.\ Ponman et al.\ 2003).

Metal distribution in the ICM can be a powerful tracer of the
history of such a gas heating in the early epoch, since the relative
timing of  metal enrichment and heating should affect the present
amount and distribution of metals in the ICM\@.  To look into this, we
assume that all galaxies synthesize a similar amount of metals per unit
stellar mass.  If metal enrichment occurred before the energy
injection,  the poor systems would carry relatively smaller
metal mass with  smaller gas mass than those in rich clusters,
 while metal abundance  would be quite similar to
those in rich clusters.
  In contrast, if metal enrichment occurred after
the energy injection,  the metal mass becomes comparable to those
in rich clusters and indicates a higher abundance.
 The difference in O and Fe distributions would constrain the timing of
 energy injection,
since O was synthesized by SN II during  galaxy star formation,
while Fe was mainly produced by SN Ia much later than this period.

The OMLR shows a larger difference in clusters and groups than the IMLR does,
which may reflect the condition that SN II products are more extended
in groups due to the effect of preheating. Fe was synthesized later by
SN Ia, and may be irrelevant to preheating.  To further investigate
the problem of heating and enrichment, we need numerical simulations
including the effect of the preheating and metal enrichment by both SN
II and SN Ia.

\begin{longtable}{lcccccl}
\caption{Comparison of IMLR, OMLR and MMLR for all systems
in B-band (\cite{kSato2008c})}
\label{tab:mlr_all}
  \hline              
\endfirsthead
  \hline
\endlastfoot
Object & IMLR & OMLR & MMLR
 & $r/r_{180}$ & $k\left<T\right>~({\rm keV})$
 & Reference \\
\hline
 \multicolumn{7}{l}{Suzaku} \\
NGC~5044
 & $2.6^{+0.2}_{-0.2} \times 10^{-3}$
 & $6.6^{+1.9}_{-1.7} \times 10^{-3}$
 & $1.6^{+0.2}_{-0.2} \times 10^{-3}$
 & $0.10$
 & $\sim 1.0$
 & this work
 \\
 & $3.6^{+0.4}_{-0.3} \times 10^{-3}$
 & $9.4^{+5.2}_{-2.1} \times 10^{-3}$
 & $2.6^{+0.4}_{-0.3} \times 10^{-3}$
 & $0.30$
 & 
 & 
 \\
Fornax
 & $4^{+0.4}_{-0.4} \times 10^{-4}$
 & $2^{+0.7}_{-0.7} \times 10^{-3}$
 & -
 & $0.13$
 & $\sim 1.3$
 & \citet{Matsushita2007a}
 \\
NGC~507
 & $6.0^{+0.4}_{-0.3} \times 10^{-4}$
 & $2.6^{+0.6}_{-0.5} \times 10^{-3}$
 & $3.7^{+0.4}_{-0.4} \times 10^{-4}$
 & $0.11$
 & $\sim 1.5$
 & \citet{kSato2008b}
 \\
 & $1.7^{+0.2}_{-0.2} \times 10^{-3}$
 & $6.6^{+3.3}_{-2.5} \times 10^{-3}$
 & $1.1^{+0.2}_{-0.2} \times 10^{-3}$
 & $0.24$
 & 
 & 
 \\
HCG~62
 & $2.0^{+0.2}_{-0.1} \times 10^{-3}$ 
 & $6.4^{+2.4}_{-4.2} \times 10^{-3}$
 & $1.0^{+0.2}_{-0.1} \times 10^{-3}$
 & $0.11$
 & $\sim 1.5$
 & \citet{Tokoi2008}
 \\
 & $4.6^{+0.7}_{-0.6} \times 10^{-3}$
 & $3.8^{+2.7}_{-3.4} \times 10^{-2}$
 & $1.5^{+0.4}_{-0.4} \times 10^{-3}$
 & $0.21$
 & 
 & 
 \\
A~262
 & $3.6^{+0.1}_{-0.1} \times 10^{-3}$
 & $1.2^{+0.3}_{-0.4} \times 10^{-2}$
 & $1.6^{+0.2}_{-0.2} \times 10^{-3}$
 & $0.10$
 & $\sim 2$
 & \citet{kSato2008c}
 \\
 & $6.7^{+0.4}_{-0.4} \times 10^{-3}$
 & $3.7^{+1.2}_{-1.2} \times 10^{-2}$
 & $2.7^{+0.7}_{-0.6} \times 10^{-3}$
 & $0.27$
 & 
 & 
 \\
A~1060
 & $5.7^{+0.4}_{-0.4} \times 10^{-3}$
 & $4.3^{+1.4}_{-1.2} \times 10^{-2}$
 & $2.4^{+0.5}_{-0.5} \times 10^{-3}$
 & $0.12$
 & $\sim 3$
 & \citet{kSato2007}
 \\
 & $4.0^{+0.4}_{-0.4} \times 10^{-3}$
 & $4.3^{+2.0}_{-1.8} \times 10^{-2}$
 & $1.6^{+0.8}_{-0.7} \times 10^{-3}$
 & $0.25$
 & 
 & 
 \\
AWM~7
 & $4.8^{+0.2}_{-0.2} \times 10^{-3}$
 & $2.6^{+0.8}_{-0.8} \times 10^{-2}$
 & $3.4^{+0.5}_{-0.5} \times 10^{-3}$
 & $0.11$
 & $\sim 3.5$
 & \citet{kSato2008a}
 \\
 & $7.6^{+0.4}_{-0.3} \times 10^{-3}$
 & $3.1^{+1.9}_{-1.2} \times 10^{-2}$
 & $6.7^{+1.1}_{-1.1} \times 10^{-3}$
 & $0.22$
 & 
 & 
 \\
 \hline
 \multicolumn{7}{l}{XMM-Newton} \\
Centaurus
 & $4^{+0.5}_{-0.5} \times 10^{-3}$
 & $3^{+0.7}_{-0.7} \times 10^{-2}$
 & -
 & $0.11$
 & $\sim 4$
 & \citet{Matsushita2007b}
 \\
\hline
\end{longtable}

\section{Summary and Conclusion}

Suzaku observations of the NGC 5044 group, with a relatively symmetric
X-ray morphology,
 allowed determination of the
 abundances of O, Mg, Si, S and Fe in the ICM
up to $0.3r_{180}$ fairly accurately.  The Fe abundance around NGC
5044 is about a solar and drops to 0.3 solar at $0.3r_{180}$. The
abundance ratios, Mg/Fe, Si/Fe and S/Fe, are close to the solar ratio,
while O/Fe is 0.5--0.6 in solar units.  
The abundance pattern indicates that most of
the Fe should have been synthesized by SN Ia.  The O distribution
closely resembles the light distribution of galaxies, in particular in
the outer region.  
 The OMLR of the NGC 5044 group is lower than those of clusters, but the
 IMLR is comparable to those of clusters.
The metal distribution
of the ICM can be used as a tracer of the past history of heating and
enrichment in the clusters of galaxies.

\end{document}